\documentclass[12pt,aps,prd,preprint,preprintnumbers,nofootinbib]{revtex4}

\usepackage[dvips]{graphicx}

\input{epsf.sty}

\def\fo{\hbox{{1}\kern-.25em\hbox{l}}}

\def\slashchar#1{\setbox0=\hbox{$#1$}           % set a box for #1
   \dimen0=\wd0                                 % and get its size 
   \setbox1=\hbox{/} \dimen1=\wd1               % get size of /
   \ifdim\dimen0>\dimen1                        % #1 is bigger 
      \rlap{\hbox to \dimen0{\hfil/\hfil}}      % so center / in box 
      #1                                        % and print #1
   \else                                        % / is bigger 
      \rlap{\hbox to \dimen1{\hfil$#1$\hfil}}   % so center #1
      /                                         % and print /
   \fi}                                         %

\def\hide#1{[hidden stuff]}

\def\beq{\begin{equation}}
\def\eeq{\end{equation}}
\def\eq{\end{equation}}
\def\to{\rightarrow}

\def\mEt{\mbox{${\hbox{$E$\kern-0.6em\lower-.1ex\hbox{/}}}_T$}\, } %missing ET

\def\bsg{\ifmmode B_d\to X_s\gamma\else $B_d\to X_s\gamma$\fi}
\def\bsglue{\ifmmode B_d\to X_s\, g\else $B_d\to X_s\, g$\fi}
\def\bstt{\ifmmode B\to X_s\tau^+\tau^-\else $B\to X_s\tau^+\tau^-$\fi}
\def\shat{\ifmmode \hat{s}\else $\hat{s}$\fi}
\def\bphik{\ifmmode B_d\to \phi K_s\else $B_d\to \phi K_s$\fi}
\def\bbarphik{\ifmmode \bar{B_d}\to \phi K_s\else $\bar{B_d}\to \phi K_s$\fi}
\def\bpsik{\ifmmode B_d\to \psi K_s\else $B_d\to \psi K_s$\fi}
\def\bbarpsik{\ifmmode \bar{B_d}\to \psi K_s\else $\bar{B_d}\to \psi K_s$\fi}
\def\bsmix{\ifmmode B_s \bar{B}_s\else $B_s \bar{B}_s$\fi}
\def\bdmix{\ifmmode B_d \bar{B}_d\else $B_d \bar{B}_d$\fi}
\def\bqmix{\ifmmode B_q \bar{B}_q\else $B_q \bar{B}_q$\fi}
\def\bclnu{\ifmmode B_d\to X_c e \bar{\nu}\else $B\to X_c e \bar{\nu}$\fi}
\def\bjpsik{\ifmmode B_d\to J/\psi~K_s\else $B_d\to J/\psi~K_s$\fi}
\def\bsee{\ifmmode B_d\to X_s e^+ e^-\else $B_d\to X_s e^+ e^-$\fi}
\def\bsmumu{\ifmmode B_d\to X_s\mu^+\mu^-\else $B_d\to X_s \mu^+\mu^-$\fi}
\def\bsll{\ifmmode B_d\to X_s\ell^+\ell^-\else $B_d\to X_s\ell^+\ell^-$\fi}
\def\bsee{\ifmmode B_d\to X_s e^+ e^-\else $B_d\to X_s e^+ e^-$\fi}

\newcommand{\newc}{\newcommand}

\newc{\asusy}{\delta a^{\rm SUSY}_\mu}
\newc{\lcal}{\int {\cal L}dt}
 
\newc{\LSP}{{\chi^0_1}}
\newc{\stauR}{{\tilde \tau_R}}
\newc{\stau}{{\tilde \tau_1}}
\newc{\mstop}{m_{\tilde{t}}}
\newc{\mHpm}{m_{H^\pm}}
\newc{\gsim}{\lower.7ex\hbox{$\;\stackrel{\textstyle>}{\sim}\;$}}
\newc{\lsim}{\lower.7ex\hbox{$\;\stackrel{\textstyle<}{\sim}\;$}}
\newc{\ie}{{\it i.e.}}          
\newc{\etal}{{\it et al.}}
\newc{\eg}{{\it e.g.}}          
\newc{\kev}{\hbox{\rm\,keV}}            
\newc{\mev}{\hbox{\rm\,MeV}}            
\newc{\gev}{\hbox{\rm\,GeV}}            
\newc{\tev}{\hbox{\rm\,TeV}}
\newc{\xpb}{\hbox{\rm\, pb}}
\newc{\xfb}{\hbox{\rm\, fb}}

\newc{\mtop}{m_t}
\newc{\mbot}{m_b}
\newc{\mz}{m_Z}
\newc{\mw}{M_W}
\newc{\alphasmz}{\alpha_s(m_Z^2)}
\newc{\swsq}{\sin^2\theta_W}
\newc{\tw}{\tan\theta_W}
\newc{\cw}{\cos\theta_W}
\newc{\sw}{\sin\theta_W}
\newc{\BR}{\hbox{\rm BR}}
\newc{\zbb}{Z\to b\bar}
\newc{\Gb}{\Gamma (Z\to b\bar b)}
\newc{\Gh}{\Gamma (Z\to \hbox{\rm hadrons})}
\newc{\rbsm}{R_b^\hbox{\rm sm}}
\newc{\rbsusy}{R_b^\hbox{\rm susy}}
\newc{\drb}{\delta R_b}

\newc{\sgn}{\mbox{sgn}}
\newc{\tbeta}{\tan\beta}
\newc{\uL}{{\tilde u_L}}
\newc{\uR}{{\tilde u_R}}
\newc{\cL}{{\tilde c_L}}
\newc{\cR}{{\tilde c_R}}
\newc{\tL}{{\tilde t_L}}
\newc{\tR}{{\tilde t_R}}
\newc{\dL}{{\tilde d_L}}
\newc{\dR}{{\tilde d_R}}
\newc{\sL}{{\tilde s_L}}
\newc{\sR}{{\tilde s_R}}
\newc{\bL}{{\tilde b_L}}
\newc{\bR}{{\tilde b_R}}
\newc{\eL}{{\tilde e_L}}
\newc{\eR}{{\tilde e_R}}
\newc{\mhp}{m_{H^\pm}}
\newc{\mhalf}{m_{1/2}}
\newc{\emt}{{e/\mu /\tau}}

\newc{\lR}{\tilde{l}_R}
\newc{\lL}{\tilde{l}_L}
\newc{\nL}{\tilde{\nu}_L}
\newc{\na}{\chi^0_1}
\newc{\nb}{\chi^0_2}
\newc{\nc}{\chi^0_3}
\newc{\nd}{\chi^0_4}
\newc{\ca}{\chi^{\pm}_1}
\newc{\cb}{\chi^{\pm}_2}
\newc{\camp}{\chi^\mp_1}
\newc{\cbmp}{\chi^\mp_1}
\newc{\capos}{\chi^{+}_1}
\newc{\caneg}{\chi^{-}_1}
\newc{\phit}{\phi_t}
\newc{\phib}{\varphi_b}
\newc{\phiew}{\phi_{ew}}
\newc{\htz}{h^0_t}
\newc{\hbz}{h^0_b}
\newc{\hewz}{h^0_{ew}}
\newc{\hsmz}{h^0_{sm}}
\newc{\huz}{h^0_u}
\newc{\hsusyz}{h^0_{susy}}

\newcommand{\drawsquare}[2]{\hbox{%
\rule{#2pt}{#1pt}\hskip-#2pt%  left vertical
\rule{#1pt}{#2pt}\hskip-#1pt%  lower horizontal
\rule[#1pt]{#1pt}{#2pt}}\rule[#1pt]{#2pt}{#2pt}\hskip-#2pt%  upper horizontal
\rule{#2pt}{#1pt}}% right vertical

\newc{\Dal}{\drawsquare{7}{0.6}}

\def\dofig#1#2{\epsfxsize=#1\centerline{\epsfbox{#2}}}
\def\dofigs#1#2#3{\centerline{\epsfxsize=#1\epsfbox{#2}%
   \hfil\epsfxsize=#1\epsfbox{#3}}}
\def\dofigsthr#1#2#3#4{\centerline{\epsfxsize=#1\epsfbox{#2}%
   \hfil\epsfxsize=#1\epsfbox{#3}%
   \hfil\epsfxsize=#1\epsfbox{#4}}}
\def\dofigsize#1#2#3#4{\centerline{\epsfxsize=#1\epsfbox{#3}%
   \hfil\epsfxsize=#2\epsfbox{#4}}}

\def\beq{\begin{equation}}
\def\eeq{\end{equation}}
\def\bea{\begin{eqnarray}}
\def\eea{\end{eqnarray}}
\catcode`@=11
\long\def\@caption#1[#2]#3{\par\addcontentsline{\csname
  ext@#1\endcsname}{#1}{\protect\numberline{\csname
  the#1\endcsname}{\ignorespaces #2}}\begingroup
    \small
    \@parboxrestore
    \@makecaption{\csname fnum@#1\endcsname}{\ignorespaces #3}\par
  \endgroup}
\catcode`@=12

\begin{document}

%\begin{titlepage}

\begin{flushright}
MSUHEP-041008 \\
hep-ph/0410181 \\
\end{flushright}

\title{
B-meson signatures of a \\
Supersymmetric U(2) flavor model}

\author{Shrihari Gopalakrishna\footnote{Current address: Dept. of Physics and Astronomy, \\ 
Northwestern University, Evanston, IL - 60208, USA.}} 
\email{shri@northwestern.edu}
\author{C.-P. Yuan} \email{yuan@pa.msu.edu}

\affiliation{
\vspace*{2mm}
{Department of Physics and Astronomy, \\ 
   Michigan State University, \\
     East Lansing, MI 48824, USA. \\ }}

\date{\today}

\begin{abstract}
We discuss B-meson signatures of a Supersymmetric U(2) flavor model, with relatively light 
(electroweak scale masses) third generation right-handed scalars. 
We impose current $B$ and $K$ meson experimental constraints on such a theory, and obtain expectations 
for \bsg, \bsglue, \bsll, \bphik, \bsmix\ mixing and the dilepton asymmetry in $B_s$. 
We show that such a theory is compatible with all current data, and furthermore, could reconcile
the apparent deviations from Standard Model predictions that have been found in some experiments.

\end{abstract}

\maketitle

%%%%%%%%%%%%%%%%%%%%%%%%%%%%%%%%%%%%%%%%%%%%%%%%%%%%%%%%%%%%%%%%%%%%%
\section{Introduction}
The Standard Model (SM) of high energy physics suffers from the gauge hierarchy problem and the flavor problem. 
The first is the fine tuning required to maintain a low electroweak mass scale ($M_{EW}$) in the theory, 
in the presence of a high scale, the Planck Scale ($M_{Pl}$). The second problem is a lack of explanation 
of the mass hierarchy and mixings of the quarks and leptons. 

Supersymmetry (SUSY) eliminates the gauge hierarchy problem by introducing for each SM particle, a new particle 
with the same mass but different spin. For example, for each SM quark/lepton a new scalar (squark/slepton),
and for each SM gauge boson a new fermion (gaugino), is introduced. If SUSY is realized in nature, the fact that we 
do not see such new particles, we believe, could be because SUSY is spontaneously broken, making the superpartners 
heavier than the mass ranges probed by experiments. Owing to a lack of understanding of how exactly SUSY is 
broken, a phenomenologically general Lagrangian, for example, the Minimal Supersymmetric Standard Model (MSSM),
is usually considered to compare with data. Various experimental searches have placed constraints on the masses and couplings in the MSSM.

Attempts have been made to address the flavor problem by proposing various flavor symmetries. In a supersymmetric 
theory, a flavor symmetry in the quark sector might imply a certain structure in the scalar sector, leading to 
definite predictions for flavor changing neutral current (FCNC) processes on which experiments have placed severe 
constraints. In the literature, a lot of attention has been devoted toward analyzing the minimal flavor violation 
(MFV) scenario, in which the scalar flavor structure is aligned with the quark sector so that the two are 
simultaneously diagonalized. In MFV, the Cabibbo-Kobayashi-Maskawa~(CKM) matrix describes the flavor changing interactions in the supersymmetric 
sector as well, and the only CP violating phase is the one in the CKM matrix. In this work, we do not assume 
such an alignment, and we will consider non-minimal flavor violation (NMFV), which we treat as a perturbation over 
the MFV case. 

In this paper we wish to explore in what form a supersymmetric extension of the SM, with a U(2) flavor symmetry, 
could influence K and B physics observables. We thus restrict ourselves to the quark and scalar-quark 
(squark) sectors. We consider an ``effective supersymmetry''~\cite{Cohen:1996vb} framework, with heavy (TeV scale) 
first two generation squarks, in order to escape neutron electric dipole moment (EDM) constraints.
This allows the possibility of having large CP violating phases in the squark sector. 
We consider a supersymmetric U(2) theory~\cite{Pomarol:1995xc,Barbieri:1995uv}, impose recent 
$K$ and $B$ meson experimental constraints and obtain predictions for \bsg, \bsglue, \bsll, \bphik, \bsmix\ 
mixing and the dilepton asymmetry in $B_s$. 
Though we consider a specific flavor symmetry, namely U(2), our conclusions 
would hold for any model with a sizable off-diagonal 32 element in the squark mass matrix.

Some $B$ physics consequences in a supersymmetric U(2) theory have been considered in 
Ref.~\cite{Barbieri:1995uv}. 
Large $\tan{\beta}$ effects in $B$ decays have been carefully analyzed in Ref.~\cite{mylgtanb}, but for 
simplicity we will restrict ourselves to the case when $\tan{\beta}$ is not too large. 
Other work along similar lines, though in more general contexts, have been presented in 
Refs.~\cite{Bertolini:1990if,Hagelin:1992tc,mygenwk,Kane:2002sp}. 
In this work we will include all dominant contributions to a particular observable in order
to include interference effects between various diagrams. This has not always been done in the literature. 
We will then study the implications of recent data from the B-factories, including 
the $b\rightarrow s$ penguin decay mode \bphik\ which shows a slight deviation from the
SM prediction. 

The paper is organized as follows: In Section~\ref{U2SUSY.SEC} we specify the supersymmetric U(2) theory 
we will work with, and the choices we make for the various SUSY and SUSY breaking parameters.
In Sections~\ref{KMIX.SEC}~and~\ref{DB2.SEC} we consider 
$\Delta S=2$ (Kaon mixing) and $\Delta B=2$ (\bdmix\ and \bsmix\ mixing) FCNC process, respectively. 
In Section~\ref{DB1.SEC} we will consider the implications of such a theory to $\Delta B=1$ FCNC processes, 
namely \bsg, \bsglue, \bsll\ and \bphik.  
We conclude in Section~\ref{CONCL.SEC}. 
We give details of various squark mixings and their diagonalization in Appendix~\ref{AppMixAng}, and collect 
loop functions that we will need in Appendix~\ref{AppLoopFcn}. 

\section{Supersymmetric U(2)}
\label{U2SUSY.SEC}
\subsection{The Model}
The supersymmetric model that we will discuss is as described in Ref.~\cite{Barbieri:1995uv}, with
the first and second generation superfields 
($\psi_a$, a=1,2) transforming as a U(2) doublet while the third generation superfield ($\psi$) is a singlet. 
The most general superpotential can be written as
\footnote{In the superpotential each term encodes the ``vertical'' gauge symmetry, which, at the weak scale, 
is $SU(3)\times SU(2)\times U(1)$. Thus ($i,j$ labels generations),
\bea
\psi_i \alpha H \psi_j \equiv \alpha_u Q_i U^c_j H_u + \alpha_u' U^c_i Q_j H_u - \alpha_d Q_i D^c_j H_d - \alpha_d' D^c_i Q_j H_d + {\rm (Lepton\ sector)}.   \nonumber
\eea}:
\beq
{\cal W} = \psi \alpha_1 H \psi + \frac{\phi^a}{M} \psi \alpha_2 H \psi_a + \frac{\phi^{ab}}{M} \psi_a \alpha_3 H \psi_b + \frac{\phi^a \phi^b}{M^2}\psi_a \alpha_4 H \psi_b + \frac{S^{ab}}{M} \psi_a \alpha_5 H \psi_b + \mu H_u H_d ,
\eeq
where $M$ is the cutoff scale below which such an effective description is valid, the $\alpha_i$ are O(1) 
constants, and three new U(2) tensor fields are introduced: $\phi^a$ a U(2) doublet, $\phi^{ab}$ a second rank 
antisymmetric U(2) tensor and $S^{ab}$ a second rank symmetric U(2) tensor. The parameter $\mu$ could be complex 
and we allow for this possibility. 
Following Ref.~\cite{Barbieri:1995uv}, we assume that U(2) is broken spontaneously by the 
vacuum expectation value (VEV)\footnote{The dynamical 
means by which this VEV is generated is left unspecified. In general, $\left<S^{22}\right>$ can be 
different from $\left< \phi^a \right>$, but for simplicity we will assume that they are the 
same. Also for simplicity, we take $\epsilon,\epsilon^\prime$ to be real.}~\cite{Barbieri:1995uv}
\bea
\left< \phi^a \right> = \pmatrix{0 \cr V};\ \ \  \left<\phi^{ab}\right> = v \epsilon^{ab};\ \ \ \left<S^{11,12,21}\right> = 0, \left<S^{22}\right> = V \ ,
\eea
with $V/M \equiv \epsilon \sim 0.02$ and $v/M \equiv \epsilon ' \sim 0.004$, in order to get the correct
quark masses. These VEV's lead to the quark mass matrix given by (we show only the 
down quark mass matrix after the $SU(2)_L$ is broken by the usual Higgs mechanism)
\bea
{\cal L} &\supset& -\pmatrix{\bar{d_R} & \bar{s_R} & \bar{b_R}} {\cal M}_d \pmatrix{d_L \cr s_L\cr b_L} + {\rm h.c.}, \\
{\cal M}_d &=&  v_d \pmatrix{O & -\lambda_1\epsilon ' & O\cr \lambda_1\epsilon '  & \lambda_2\epsilon & \lambda_4\epsilon\cr  O & \lambda_4'\epsilon & \lambda_3}, \nonumber
\eea
where $v_d = \left< h_d \right>$ is the VEV of the Higgs field. In ${\cal M}_d$, the $\lambda_i$'s are O(1) 
(complex) coefficients, given in terms of the $\alpha_i$'s. Ref.~\cite{Barbieri:1995uv} shows that such a 
pattern of the mass matrix explains the quark masses and CKM elements.

If U(2) is still a good symmetry at the SUSY breaking scale, and broken (spontaneously) only below the 
SUSY breaking scale, the SUSY breaking terms would have a structure dictated by U(2). 
For our purposes it is sufficient to consider the down sector squark mass matrices, and they are given as
\bea
{\cal L} \supset -\pmatrix{\tilde d_L^* & \tilde s_L^* & \tilde b_L^*} {\cal M}^2_{LL} \pmatrix{\tilde d_L\cr \tilde s_L\cr \tilde b_L} - \pmatrix{\tilde d_R^* & \tilde s_R^* & \tilde b_R^*} {\cal M}^2_{RR} \pmatrix{\tilde d_R\cr \tilde s_R\cr \tilde b_R} \nonumber \\
+ \left( \pmatrix{\tilde d_R^* & \tilde s_R^* & \tilde b_R^*} {\cal M}^2_{RL} \pmatrix{\tilde d_L\cr \tilde s_L\cr \tilde b_L} + {\rm h.c.} \right),
\eea
\bea
{\cal M}^2_{LL} &=& {\cal M}_d^\dag {\cal M}_d + \pmatrix{m_1^2 & i\epsilon^\prime m_5^2 & 0\cr
	 -i\epsilon^\prime m_5^2 & m_1^2+\epsilon^2 m_2^2 & \epsilon m_4^{2*}\cr 
	0 & \epsilon m_4^2 & m_3^2}_{LL} + {\rm D-term}, \nonumber \\ 
{\cal M}^2_{RR} &=& {\cal M}_d {\cal M}_d^\dag + \pmatrix{m_1^2 & i\epsilon^\prime m_5^2 & 0\cr
	 -i\epsilon^\prime m_5^2 & m_1^2+\epsilon^2 m_2^2 & \epsilon m_4^{2*}\cr 
	0 & \epsilon m_4^2 & m_3^2}_{RR} + {\rm D-term}, \nonumber \\
{\cal M}^2_{RL} &=& \mu^* \tan{\beta} {\cal M}_d + v_d \pmatrix{O & -A_1 \epsilon ' & O\cr A_1 \epsilon '  & A_2\epsilon & A_4 \epsilon\cr  O & A_4'\epsilon & A_3},
\label{MSQ.EQ}
\eea
where $m_i^2$ and $A_i$ are determined by the SUSY breaking mechanism. 
Here $m_1^2,m_2^2,m_3^2~{\rm and}~m_5^2$ are real, while $m_4^2~{\rm and}~A_i$ could be complex. 
We will assume that the $A_i$ are of order $A$, a common mass scale. The D terms are flavor diagonal, 
and since we are interested in FCNC processes, we will not write them 
in detail, but will think of them as included in $m_1^2$ and $m_3^2$.

Thus far we have presented the mass matrices in the gauge basis. In the following sections, we will 
work in the superKM basis in which the quark mass matrix is diagonal, and the quark field rotations 
that diagonalize the quark mass matrix are applied to the squarks, whose mass matrix would also 
have been diagonalized in the MFV scheme. Since we will not assume an MFV structure, in the 
superKM basis, there would be small off-diagonal terms in the squark mass matrix, which we treat as 
perturbations. The structure of the squark mass matrix in the superKM basis 
is similar to that in Eq.~(\ref{MSQ.EQ}) owing to the smallness of the mixing angles that 
diagonalize the quark 
mass matrix.

\subsection{SUSY parameters}
\label{SUSYPARAM.SEC}
Lacking specific knowledge about the SUSY breaking mechanism realized in nature, 
we make some assumptions on the SUSY mass spectrum. Neutron EDM places strong constraints 
on the CP violating phases and the masses of the first two generations of scalars. 
To satisfy this and other collider constraints, we consider an ``effective SUSY'' 
framework in which the scalars of the first two generations are heavy, 
suppressing EDM, and allowing for larger CP violating phases. 
Defining the scalar mass scale, $m_0 \sim 1~{\rm TeV}$, we take all $m_i \sim m_0$ 
except for $m_{\tilde t_R,\tilde b_R} \equiv {m_3}_{RR} \sim 100~{\rm GeV}$. We take $A \sim m_0$, 
the gaugino mass parameter $M_2$ and charged-Higgs masses to be $250~{\rm GeV}$ and the 
gluino mass to be $300~{\rm GeV}$\footnote{The Tevatron bounds on the stop, sbottom and 
gluino masses are discussed in Ref.~\cite{Affolder:1999wp}. 
We note here that the bounds in general get less stringent as the 
neutralino mass increases.}. We assume such a spectrum just above the weak scale without 
specifying what mechanism of SUSY breaking and mediation might actually give rise to it. 
As we will show later, if realized in nature such a spectrum would lead to enhancements in the 
processes we are considering here.

The rates of various FCNC processes follow from the mass matrix that we have specified 
in Eq.~(\ref{MSQ.EQ}). We will work in the superKM basis. 
The interaction vertices in the mass basis are obtained by diagonalizing the mass matrices in 
Eq.~(\ref{MSQ.EQ}), and the perturbative diagonalization to leading order is shown in 
Appendix.~\ref{AppMixAng}.

The dominant NMFV SUSY contributions to FCNC processes would be due to 
the 32 and 23 entries in Eq.~(\ref{MSQ.EQ}), since they are the biggest off-diagonal terms. 
For convenience we define
\beq
\delta_{32,23}^{RL,RR,LL} \equiv \frac{({\cal M}^2_{RL,RR,LL})_{32,23}}{m_0^2}.
\eeq
Since we have written down an effective theory and not specified the dynamics of U(2) and 
SUSY breaking, we can only specify the order of magnitude of $\delta_{32,23}$. 
To parametrize this uncertainty we write, 
\beq
\delta_{32,23}^{RL} = \frac{v_d A\epsilon}{m_0^2}\, d_{32,23}^{RL} , \qquad
\delta_{32,23}^{LL,RR} = \frac{\epsilon m_4^2}{m_0^2}\, d_{32,23}^{LL,RR}  \ ,
\label{DELDEF.EQ}
\eeq
where we have denoted the unknown $\mathcal{O}(1)$ coefficients by $d_{32,23}^{LL,RR,RL}$. 
 
\begin{table}[h]
\centering
\begin{tabular}{||c|c||c|c||}
\hline
$m_0$ & $1000\ $ & $\tan{\beta}$ & $5\ $ \cr 
${m_{\tilde b_R,\tilde t_R}}$ & $100\ $ & $\mu$ & $200\, e^{i\,2.2}\ $ \cr 
${m_{\tilde d_R,\tilde s_R}}$ & $1000\ $ & $M_2$ & $250\ $ \cr 
${m_{\tilde q_L}}$ & $1000\ $ & $M_{\tilde g}$ & $300\ $ \cr
$A$ & $1000\ $ & $m_{H^\pm}$ & $250\ $ \cr
\hline
%\end{tabular}
%\begin{tabular}{||c|c||c|c||}
\hline
$\ d_{32}^{RL}\ $ & $~2\,e^{i\,3.2}\ $ & $\ d_{32}^{RR}\ $ & $1.75\,e^{i\,1.6}\ $ \cr 
\hline
\end{tabular}
\caption{Default SUSY parameters for this work that satisfy all 
experimental constraints discussed in this paper. All masses are in GeV.}
\label{SPARAM.TAB}
\end{table}
\noindent We summarize our choice of the parameters in Table~\ref{SPARAM.TAB}.
For these values, from Eq.~(\ref{DELDEF.EQ}), the natural sizes of $\delta_{32,23}$ are given by
\beq
\delta_{32,23}^{RL} = 6.82\times 10^{-4}\, d_{32,23}^{RL} \ , \qquad 
\delta_{32,23}^{LL,RR} = 0.02\, d_{32}^{LL,RR} \ .
\label{DELNUM.EQ}
\eeq

\begin{figure}
\dofigs{2.5in}{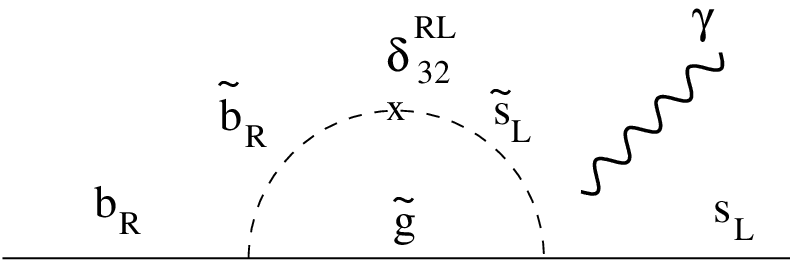}{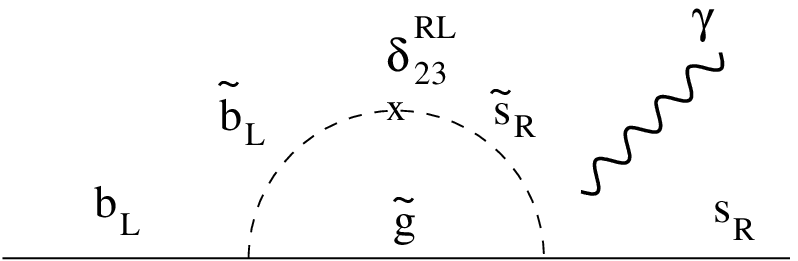}
\caption{Gluino contribution to \bsg. The diagram on the left, proportional to $\delta_{32}^{RL}$, 
has the lighter scalar $\tilde b_R$, while the one on the right, proportional to 
$\delta_{23}^{RL}$, only has heavier scalars and is therefore relatively suppressed.}
\label{BSG3223.FIG}
\end{figure}

We will find in the rest of this paper that $\delta^{RL}$ induces NMFV $\Delta B=1$ FCNC 
processes dominantly, while $\delta^{RR,LL}$ induces $\Delta S=2$ and $\Delta B=2$ FCNC processes.
Though the $\delta_{32}^{RL}$ and $\delta_{23}^{RL}$ elements have similar magnitudes, 
the $\delta_{32}^{RL}$ gluino NMFV contribution to $\Delta B=1$ FCNC processes is 
larger, since we take $\tilde b_R$ to be much lighter than the other scalars, and the 
$\delta_{23}^{RL}$ gluino diagrams are relatively suppressed by the heavier 
$\tilde b_L$ mass. Therefore, in this work
we will include only the dominant $\delta_{32}^{RL}$ contribution.
We illustrate this in Fig.~\ref{BSG3223.FIG}, where we show the gluino contribution to \bsg\ as an example. 
Similarly, owing to the smaller $\tilde b_R$ mass, the $\delta_{32,23}^{RR}$ 
NMFV contribution to $\Delta S=2$ and $\Delta B=2$ FCNC processes is 
relatively larger compared to the $\delta_{32,23}^{LL}$ contribution. 
We note here that, from Eq.~(\ref{THDIAGSB.EQ}) in Appendix~\ref{AppMixAng}, the sbottom mixing 
angle is negligibly small, and therefore, we ignore sbottom mixing effects; stop mixing is not 
as small and we include its effects.

In the next three sections we will discuss the implication of the U(2) model to 
$\Delta S=2$, $\Delta B=2$ and $\Delta B=1$ FCNC processes. From this we will see that 
present experimental data are compatible with the values shown in Table~\ref{SPARAM.TAB}, 
and we will obtain expectations for some measurements that are forthcoming.
We will present plots of different FCNC effects by varying a couple of parameters at a time,  
while keeping all others fixed at the values shown in Table~\ref{SPARAM.TAB}.

\section{$\Delta S=2$ FCNC process}
\label{KMIX.SEC}
The CP violation parameter $\epsilon_K$ due to mixing in the Kaon sector has been 
measured to be~\cite{Eidelman:2004wy}
\beq
|\epsilon_K| = \left( 2.284\pm 0.014 \right) \times 10^{-3} \ .
\eeq  
We wish to estimate the new physics contributions to $\epsilon_K$ in the scenario 
that we are considering.
Here we note that even though the direct CP violation parameter $\epsilon_K^\prime/\epsilon_K$ has 
also been measured, large hadronic uncertainties do not permit us to constrain new physics models 
through this observable.

Kaon mixing is governed by the $\Delta S=2$ effective Hamiltonian
\beq
{\cal H}^{eff}_{\Delta S=2} = \sum_{i=1}^{5} C_i Q_i + \sum_{i=1}^{3} \tilde{C}_i \tilde{Q}_i \ ,
\eeq
where, 
\bea
Q_1 &=& \bar{d}_L^\alpha \gamma_\mu s_L^\alpha \bar{d}_L^\beta \gamma^\mu s_L^\beta \ , \nonumber \\
Q_2 &=& \bar{d}_R^\alpha s_L^\alpha \bar{d}_R^\beta s_L^\beta \ , \nonumber \\
Q_3 &=& \bar{d}_R^\alpha s_L^\beta \bar{d}_R^\beta s_L^\alpha \ , \nonumber \\
Q_4 &=& \bar{d}_R^\alpha s_L^\alpha \bar{d}_L^\beta s_R^\beta \ , \nonumber \\
Q_5 &=& \bar{d}_R^\alpha s_L^\beta \bar{d}_L^\beta s_R^\alpha \ .
\label{DS2OPS.EQ}
\eea
The operators $\tilde{Q}_i$ (i=1,2,3) are obtained by exchanging $L\leftrightarrow R$.
In the SM and the new physics model we are considering, the dominant 
contributions are to $Q_1$, as we explain later in this section.  
The CP violation parameter $\epsilon_K$ is then given by (see for example Ref.~\cite{Branco:1994eb})
\bea
\epsilon_K = e^{i \pi/4}\frac{1}{3\sqrt{2}}\frac{m_K B_K f_K^2}{\Delta m_K} ~ {\rm Im}\left[C_1(m_K)\right] \ ,
\eea
where $B_K$ is the Bag parameter and $f_K$ is the Kaon decay constant. 
 
In addition to the SM $W$ box diagram contribution to $C_1$, in the supersymmetric U(2) 
theory we are considering, the charged-Higgs and chargino MFV contributions could be sizable.
The dominant MFV contributions to $C_1$ can be written as
\beq
C_1^{MFV} = C_1^{W} + C_1^H + C_1^\chi  \ , 
\eeq
which is the sum of the SM $W$, the charged-Higgs, and the chargino contributions, respectively. \\
{\bf SM contribution}: The SM $W$ contribution is~\cite{Buchalla:1995vs}
\bea
C_1^{SM}(m_t) = C_1^{W}(m_t) = \frac{G_F^2 m_W^2}{4\pi^2} 
~~ \Biggl{\{} \left( V_{td}^* V_{ts} \right)^2 ~ S_0(x_t) \nonumber \\ 
\Biggl. + \left(V_{cd}^*V_{cs}\right)^2 ~ S_0(x_c)
+ 2 \left(V_{td}^*V_{ts} V_{cd}^*V_{cs} \right) ~ S_0(x_t,x_c)  \Biggl{\}} \ ,
\label{DB2C1WMT.EQ}
\eea 
where the function $S_0$ is given in Appendix~\ref{AppLoopFcn}, Eq.~(\ref{BOXSMS0.EQ}), and
$x_t\equiv m_t^2/m_W^2$, $x_c\equiv m_c^2/m_W^2$. The QCD correction due to renormalization group
running from $m_t$ to $m_b$ gives
\bea
C_1^{SM}(m_K) = C_1^{W}(m_K) = \frac{G_F^2 m_W^2}{4\pi^2} 
~~ \Biggl{\{} \left( V_{td}^* V_{ts} \right)^2 \eta_{K33} ~ S_0(x_t) \nonumber \\ 
\Biggl. + \left(V_{cd}^*V_{cs}\right)^2 \eta_{K22} ~ S_0(x_c)
+ 2 \left(V_{td}^*V_{ts} V_{cd}^*V_{cs} \right) \eta_{K32} ~ S_0(x_t,x_c)  \Biggl{\}} \ ,
\eea
where the $\eta_K$ are QCD correction factors given in Eq.~(\ref{PARAMK.EQ}) below, 
and $V_{ij}$ are the CKM matrix elements. \\
{\bf Charged-Higgs contribution}: Supersymmetric theories require two Higgs doublets to give masses to the 
up and down type fermions. The Higgs doublets contain the charged-Higgs $H^\pm$, and
the dominant charged-Higgs-top contribution is~\cite{Branco:1994eb}\footnote{The charged-Higgs also
contributes to the operator $\tilde Q_2$, which becomes important only at large $\tan{\beta}$.}
\bea
C_1^H(m_t) &=&  \frac{G_F^2 m_W^2}{4\pi^2} \left( V_{td}^* V_{ts} \right)^2 ~\left[-F_V^H\right] \ , \label{DB2C1HMT.EQ} \\
F_V^H &=& \frac{1}{4\tan^4\beta}x_t^2~Y_1(r_H,r_H,x_t,x_t) + \frac{1}{2\tan^2\beta}x_t^2~Y_1(1,r_H,x_t,x_t) - \frac{2}{\tan^2\beta}x_t~Y_2(1,r_H,x_t,x_t) \ , \nonumber
\eea
where $r_H\equiv m_H^2/m_W^2$, and the functions $Y_1$ and $Y_2$ are given in 
Appendix~\ref{AppLoopFcn}, Eq.~(\ref{BOXHY.EQ}). \\
{\bf Chargino contribution}: The dominant chargino-right-handed-stop contribution is~\cite{Branco:1994eb}
\bea
C_1^\chi(m_t) &=&  \frac{G_F^2 m_W^2}{4\pi^2} \left( V_{td}^* V_{ts} \right)^2 \left[ - F_V^\chi \right] \ , \label{DB2C1CHMT.EQ}  \\
F_V^\chi &=& \frac{1}{4} |\Gamma_{\chi R}^{(i)}|^2 |\Gamma_{\chi R}^{(j)}|^2 ~ Y_1(r_{\tilde t_2},r_{\tilde t_2},s_i,s_j) \ ,  \nonumber
\eea
where $r_{\tilde t_2}\equiv m_{\tilde t_2}^2/m_W^2$, $s_{1,2}\equiv m_{\tilde \chi_{1,2}}^2/m_W^2 $, 
and  the coupling is given by 
\beq
\Gamma_{\chi R}^{(i)} = \sqrt{2}({\cal C}_R^*)_{1i}({\cal C}_{\tilde t}^*)_{12} - \frac{({\cal C}_R^*)_{2i} ({\cal C}_{\tilde t}^*)_{22}  }{\sin\beta} \frac{m_t}{m_W} \ ,
\eeq
with the chargino and stop diagonalization matrices $({\cal C}_R)$ and $({\cal C}_{\tilde t})$ given in 
Appendix~\ref{AppMixAng}, Eqs.~(\ref{CDIAGCH.EQ})~and~(\ref{CDIAGSB.EQ}), respectively . 
Taking into account renormalization group running, we have
\beq
C_1^{H,\chi}(m_K) \approx  \eta_{K33}~C_1^{H,\chi}(m_t) \ .
\eeq
{\bf Gluino contribution}: In general, the NMFV gluino contributions induce many operators shown in
Eq.~(\ref{DS2OPS.EQ}), but in the model we are considering, these are not significant due to a 
suppression from the heavy 
$\tilde d$ and $\tilde s$ masses, Glashow-Iliopoulos-Maiani~(GIM) suppression 
owing to their approximate degeneracy (split only
by $\mathcal{O}(\epsilon^2)$, cf. Eq.~(\ref{MSQ.EQ})), and the contribution from the relatively light 
right-handed sbottom being suppressed by its small mixing to the first two generations. 
Moreover, owing to the structure of the mass matrix, Eq.~(\ref{MSQ.EQ}), the gluino contribution is 
real, and hence does not contribute to $\epsilon_K$.

In our numerical analysis, we take the following values for the various 
parameters~\cite{Eidelman:2004wy,Buchalla:1995vs}:
\bea
\eta_{K33} &=& 0.57 \ , \quad \eta_{K22} = 1.38 \ , \quad \eta_{K32} = 0.47 \ , \nonumber \\
f_K &=& 0.160 ~{\rm GeV} \ , \quad 0.6 < B_K < 0.9 \ ,  \label{PARAMK.EQ} \\
m_K &=& 0.497 ~{\rm GeV} \ , \quad \Delta m_K = (3.48\pm 0.01)\times 10^{-15} ~{\rm GeV} \ , 
\quad m_c=(1.2\pm 0.2)~{\rm GeV} \ . \nonumber 
\eea
   
\begin{figure}
\dofigs{2.5in}{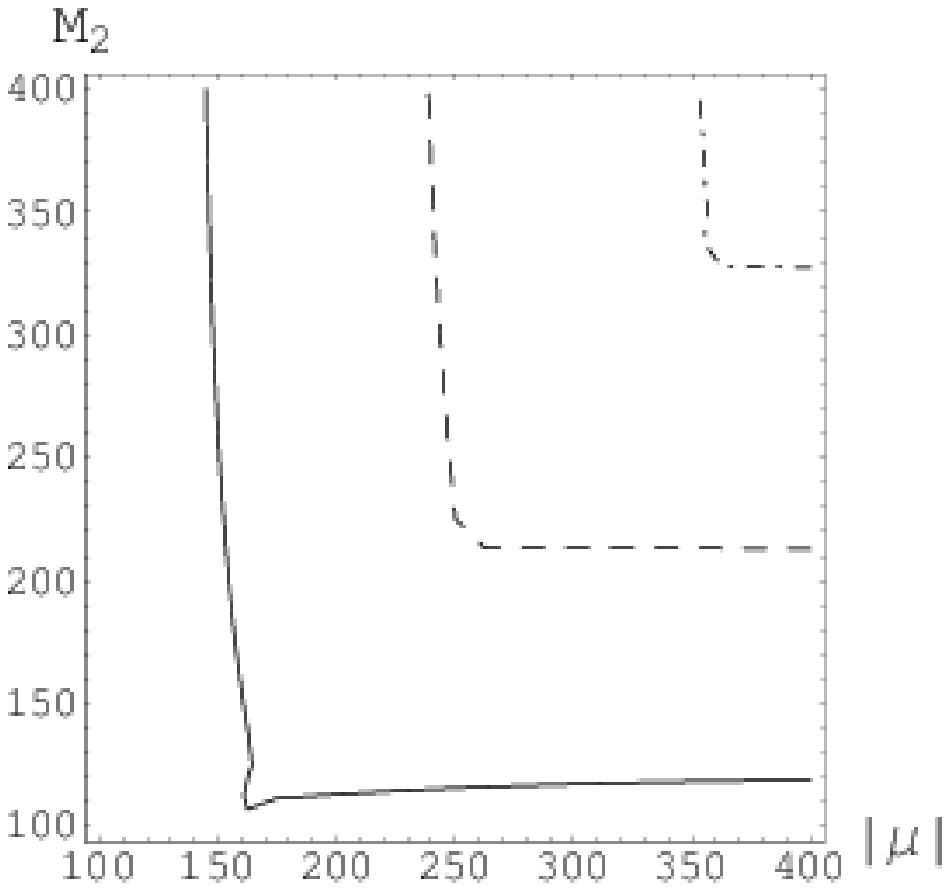}{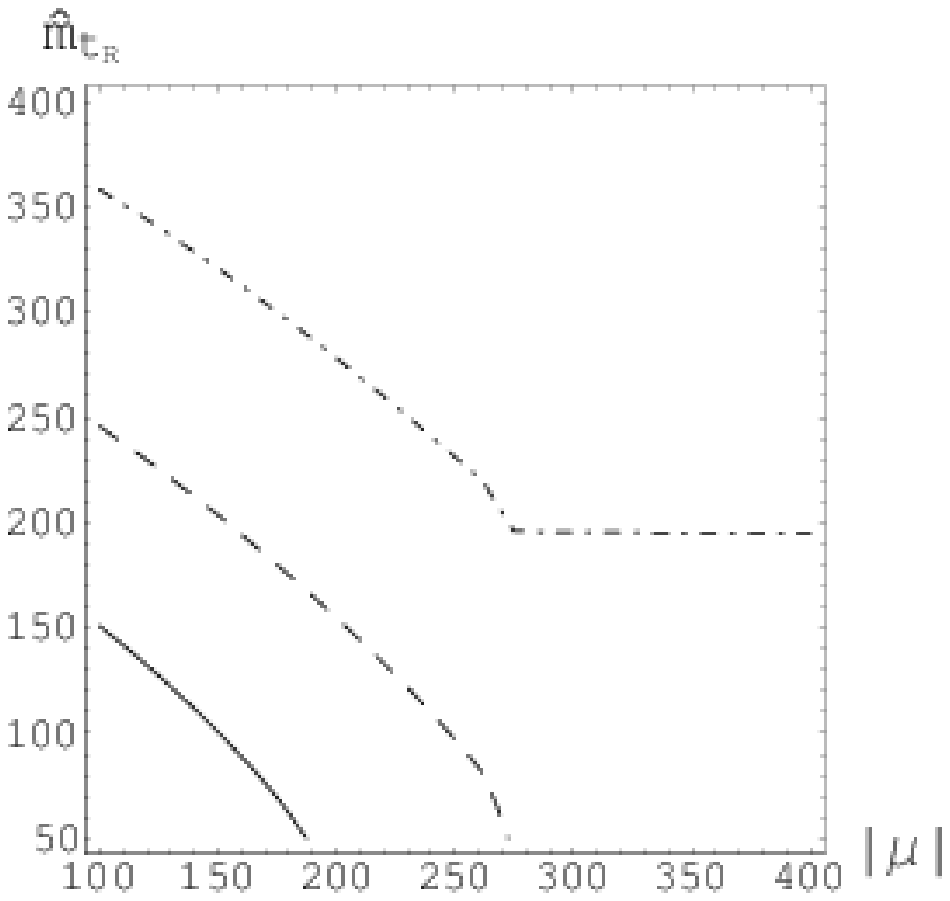}
\caption{The (dash-dot,~dash,~solid) curves are (1.05,~1.1,~1.2) contours of 
${\rm Im}(C_1^{MFV})/{\rm Im}(C_1^{SM})$, showing the MFV contributions to Kaon mixing relative to the SM.
Parameters not shown on a plot's axes are fixed as shown in Table~\ref{SPARAM.TAB}.
}
\label{EPSKMFV.FIG}
\end{figure}

The SM prediction for $\epsilon_K$ is in agreement with the experimental data, but it should be noted
that there is considerable uncertainty in the lattice computation of the Bag parameter $B_K$ 
(see Eq.~(\ref{PARAMK.EQ})). The chargino and charged-Higgs contributions to $C_1$ add 
constructively with the SM contribution. 
Therefore, if the true value of $B_K$ is taken to be closer to the lower limit, 
we can allow MFV contributions to be up by a factor of 1.2 compared to the SM value; 
i.e., ${\rm Im}(C_1^{MFV})/{\rm Im}(C_1^{SM}) \lsim 1.2$. Fig.~\ref{EPSKMFV.FIG} shows 
the region of MFV parameter space where this is satisfied. This justifies some of the choices we make in the
list shown in Table~\ref{SPARAM.TAB}. 

\section{$\Delta B=2$ FCNC processes}
\label{DB2.SEC}
\subsection{General formalism}
We start by discussing in general \bqmix\ mixing and later specialize in succession to \bdmix\ (q=d) and
to \bsmix\ (q=s). The $\Delta B=2$ effective Hamiltonian is given by~\cite{Becirevic:2001jj}:
\beq
{\cal H}^{eff}_{\Delta B=2} = \sum_{i=1}^{5} C_i Q_i + \sum_{i=1}^{3} \tilde{C}_i \tilde{Q}_i \ ,
\eeq
\noindent where, for $B_q$,
\bea
Q_1 &=& \bar{q}_L^\alpha \gamma_\mu b_L^\alpha \bar{q}_L^\beta \gamma^\mu b_L^\beta \ , \nonumber \\
Q_2 &=& \bar{q}_R^\alpha b_L^\alpha \bar{q}_R^\beta b_L^\beta \ , \nonumber \\
Q_3 &=& \bar{q}_R^\alpha b_L^\beta \bar{q}_R^\beta b_L^\alpha \ , \nonumber \\
Q_4 &=& \bar{q}_R^\alpha b_L^\alpha \bar{q}_L^\beta b_R^\beta \ , \nonumber \\
Q_5 &=& \bar{q}_R^\alpha b_L^\beta \bar{q}_L^\beta b_R^\alpha \ .
\eea
The operators $\tilde{Q}_i$ (i=1,2,3) are obtained by exchanging $L\leftrightarrow R$. 
The Wilson coefficients $C_i$ are run down from the SUSY scale, $M_S$, 
using~\cite{Becirevic:2001jj}
\beq
C_r(m_b)=\sum_{i} \sum_{s} \left(b_i^{(r,s)}+\eta c_i^{(r,s)}\right)\eta^{a_i} C_s(M_S)
\label{BSEVOL.EQ}
\eeq
\noindent where $\eta \equiv \alpha_s(M_S)/\alpha_s(m_t)$ and the $a_i$, $b_i$ and $c_i$ are constants given 
in Ref.~\cite{Becirevic:2001jj}.

\noindent The matrix elements of the $Q_i$ in the vacuum insertion approximation are given by~\cite{Becirevic:2001jj,Ko:2002ee}. 
\bea
\left<\bar{B}_q|Q_1(\mu)|B_q\right> &=& \frac{2}{3}m_{B_q}^2 f_{B_q}^2 B_1(\mu) \ , \nonumber \\
\left<\bar{B}_q|Q_2(\mu)|B_q\right> &=& -\frac{5}{12}\left( \frac{m_{B_q}}{m_b+m_q}\right)^2 m_{B_q}^2 f_{B_q}^2 B_2(\mu) \ ,  \nonumber \\ 
\left<\bar{B}_q|Q_3(\mu)|B_q\right> &=& \frac{1}{12}\left( \frac{m_{B_q}}{m_b+m_q}\right)^2 m_{B_q}^2 f_{B_q}^2 B_3(\mu) \ ,  \nonumber \\
\left<\bar{B}_q|Q_4(\mu)|B_q\right> &=& \frac{1}{2}\left( \frac{m_{B_q}}{m_b+m_q}\right)^2 m_{B_q}^2 f_{B_q}^2 B_4(\mu) \ ,  \nonumber \\
\left<\bar{B}_q|Q_5(\mu)|B_q\right> &=& \frac{1}{6}\left( \frac{m_{B_q}}{m_b+m_q}\right)^2 m_{B_q}^2 f_{B_q}^2 B_5(\mu) \ ,
\label{BBMATEL.EQ}
\eea
where we take for the decay constants $f_{B_q} = 0.2\pm 0.03$~GeV and the Bag parameters
(at scale $m_b$) $B_1=0.87$, $B_2=0.82$, $B_3=1.02$, $B_4=1.16$ and $B_5=1.91$~\cite{Becirevic:2001jj,Ko:2002ee}. 

The $B_q$ mass difference is given by
\bea
\Delta m_{B_q} = 2 \left| M_{12}(B_q) \right| \ , 
\eea
where $M_{12}(B_q)$ is the off-diagonal Hamiltonian element for the \bqmix\ system, 
and is given by
\bea
M_{12} &=& M_{12}^{SM} + M_{12}^{SUSY} \ ,\nonumber \\
\Gamma_{12} &\approx& \Gamma_{12}^{SM} \ . \nonumber 
\eea
$\Gamma_{12}$ to an excellent approximation is dominated by the SM tree decay modes. 
From Refs.~\cite{Buras:1997fb,Ko:2002ee} we have,
\bea
|M_{12}(B_q)| = \frac{1}{2m_{B_q}} \left| \left<B_q|{\cal H}^{eff}_{\Delta B=2}|\bar{B}_q\right> \right| \ , 
\label{M12BQ.EQ}  \nonumber  \\
\Gamma_{12}^{SM} = (-1)\frac{G_F^2 m_b^2 m_{B_q} B_{B_q} f_{B_q}^2 }{8\pi} \left[ v_t^2+\frac{8}{3}v_c v_t (z_c + \frac{1}{4}z_c^2-\frac{1}{2}z_c^3) \right. \nonumber \\
\left. + v_c^2\left( \sqrt{1-4z_c}(1-\frac{2}{3}z_c)+\frac{8}{3}z_c+\frac{2}{3}z_c^2-\frac{4}{3}z_c^3-1 \right) \right] \ ,
\label{GAM12BQ.EQ}
\eea
where $v_x\equiv V_{xb}V_{xq}^*$, $z_c\equiv m_c^2/m_b^2$ and we take 
$B_{B_q} \approx 1.37$.

The dilepton asymmetry in $B_q$ is given by~\cite{Randall:1998te}
\beq
A_{ll}^{B_q} \equiv \frac{N(B_q B_q) - N(\bar{B}_q \bar{B}_q)}{N(B_q B_q) + N(\bar{B}_q \bar{B}_q)} = {\rm Im}\left({\frac{\Gamma_{12}}{M_{12}}}\right) \ .
\eeq

We discuss next the SM and new physics contributions to the coefficients $C_i$ and $\tilde C_i$. \\
{\bf MFV contribution}: 
The SM $W$ contribution is almost identical to that shown in Eq.~(\ref{DB2C1WMT.EQ}) but for 
the fact that it is sufficient to keep only the top contribution (the $S_0(x_t)$ term) and 
changing the CKM factor to $\left( V_{tq}^* V_{tb} \right)^2$. The new physics MFV charged-Higgs 
and chargino contributions are again identical to 
Eqs.~(\ref{DB2C1HMT.EQ})~and~(\ref{DB2C1CHMT.EQ}), respectively, with the same change for 
the CKM factors. $C_1(m_t)$ is evolved down to $m_b$ using Eq.~(\ref{BSEVOL.EQ}).

{\bf Gluino contribution}: We only include the dominant gluino-right-handed-sbottom box diagrams 
with $\delta_{32}^{RL}$ and $\delta_{32}^{RR}$ mass insertions, since $\tilde b_R$ is the only
relatively light down type squark in our scenario. These contributions are given by~\cite{Hagelin:1992tc}
\bea
\tilde{C}_1^{\tilde g}(M_{\tilde{g}}) &=& i g_S^4 \left[\Gamma_{b3}^{RR}\left(\Gamma_{q3}^{RR}\right)^* \right]^2 \left[\frac{1}{36} \tilde{I}_4 + \frac{1}{9} M_{\tilde{g}}^2  I_4 \right] \ ,  \nonumber \\
\tilde{C}_2^{\tilde g}(M_{\tilde{g}}) &=& i g_S^4 \left[\Gamma_{b3}^{RL}\left(\Gamma_{q3}^{RL}\right)^* \right]^2 \left[ \frac{3}{2} M_{\tilde{g}}^2 I_4 \right] \ , \nonumber \\
\tilde{C}_3^{\tilde g}(M_{\tilde{g}}) &=& -i g_S^4 \left[\Gamma_{b3}^{RL}\left(\Gamma_{q3}^{RL}\right)^* \right]^2 \left[ \frac{1}{2} M_{\tilde{g}}^2 I_4 \right] \ ,
\label{C1BQGL.EQ}
\eea 
with the box integrals $I_4$ and $\tilde I_4$ given in Appendix~\ref{AppLoopFcn}. The couplings 
are given by 
\bea
\Gamma_{b3}^{RR} = \cos{\theta_{32}^{RR}} &,& \quad   \Gamma_{b3}^{RL} = \cos{\theta_{32}^{RL}} \ , \nonumber  \\
\Gamma_{d3}^{RR} = \sin{\theta_{12}^{RR}} \sin{\theta_{32}^{RR}} e^{-i(\gamma_{32}^{RR}+\gamma_{12}^{RR})} \ &,& \quad
\Gamma_{s3}^{RR} = - \cos{\theta_{12}^{RR}} \sin{\theta_{32}^{RR}} e^{-i\gamma_{32}^{RR}} \ , \nonumber  \\
\Gamma_{d3}^{RL} = \sin{\theta_{12}^{RL}} \sin{\theta_{32}^{RL}} e^{-i(\gamma_{32}^{RL}+\gamma_{12}^{RL})} \ &,& \quad
\Gamma_{s3}^{RL} = - \cos{\theta_{12}^{RL}} \sin{\theta_{32}^{RL}} e^{-i\gamma_{32}^{RL}} \ ,
\label{GQ3COUP.EQ}
\eea
obtained from the $3\times 3$ mixing matrix that is the product of ${\mathcal C}_{\tilde d_{R} \tilde s_{R}}$
and ${\mathcal C}_{\tilde b_R \tilde s_R}$, 
with the mixing angles $\theta$ and phases $\gamma$ given in Appendix~\ref{AppMixAng}. In our U(2) model, 
if $m_4$ is of the same order as $A$, based on the estimate in Eq.~(\ref{DELNUM.EQ}), we expect $\tilde{C}_1^{\tilde g}$ 
to receive the dominant gluino contribution from $\delta_{32}^{RR}$. 
We will focus on this contribution in the following. 

We point out in Appendix~\ref{AppMixAng}, Eq.~(\ref{MDSLLRR.EQ}), that 
$\tilde d_{R}\tilde s_{R}$ mixing can be generically large (near maximal),
in which case the gluino contributions to both \bdmix\ and \bsmix\ mixing can be sizable.
However, if $\epsilon^\prime m_5^2 \ll \epsilon^2 m_2^2$, this mixing can be small and the gluino
contribution to \bdmix\ mixing is negligible since it is proportional to $\sin{\theta_{12}^{RR}}$, 
cf. Eqs.~(\ref{C1BQGL.EQ})~and~(\ref{GQ3COUP.EQ}). The gluino contribution to \bsmix, however, can
still be sizable in either case since it is proportional to $\cos{\theta_{12}^{RR}}$.

\subsection{\bdmix\ mixing}
\label{BDMIX.SEC}
The \bdmix\ mass difference ($\Delta m_d$), and CP violation in \bpsik\ ($a_{\psi K_s}$) have been measured to 
be~\cite{Eidelman:2004wy,HFAG},
\bea
\Delta m_d &=& 0.502 \pm 0.007 ~ {\rm ps^{-1}} \ , \nonumber \\
a_{\psi K_s} &=& 0.725\pm 0.037 \ .
\label{S2BEXP.EQ}
\eea
In the SM, the usual notation is, $a_{\psi K_s}^{SM} \equiv \sin{2\beta}$.

\begin{figure}
\dofigs{2.5in}{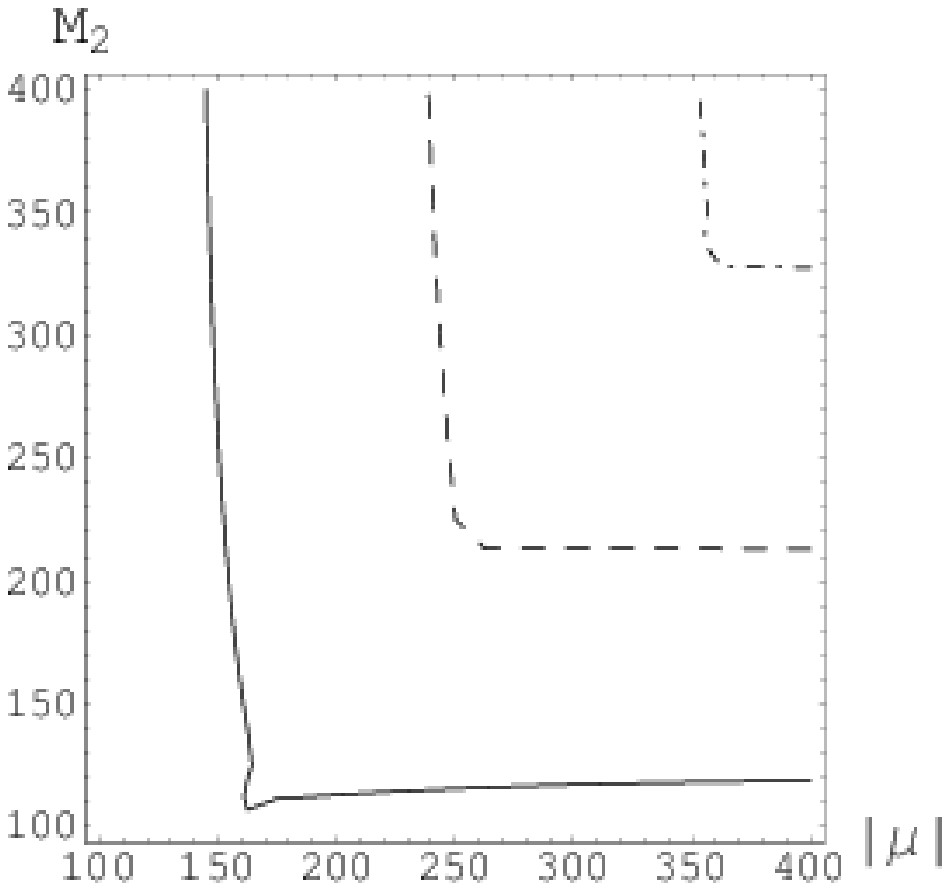}{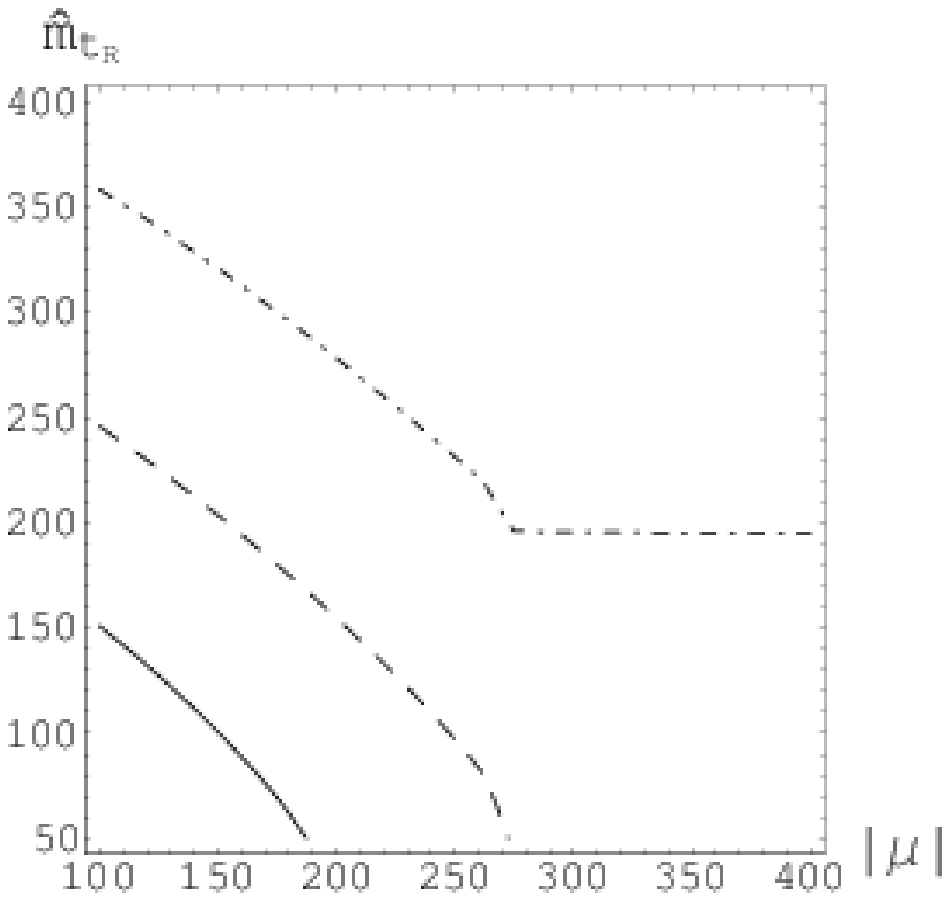}
\caption{The (dash-dot,~dash,~solid) curves are (1.1,~1.2,~1.3) contours of $|C_1^{MFV}/C_1^{SM}|$,
showing the MFV contributions to \bdmix\ mixing relative to the SM.
Parameters not shown on a plot's axes are fixed as shown in Table~\ref{SPARAM.TAB}.
}
\label{DMBDMFV.FIG}
\end{figure}

As we have already pointed out in Section~\ref{KMIX.SEC}, the charged-Higgs and chargino MFV contributions
add constructively with the SM contribution. The SM prediction agrees quite well with the data, but
given the uncertainty in $f_{B_d}$, cf. below Eq.~(\ref{BBMATEL.EQ}), it might be possible to 
accommodate an MFV contribution up to a factor of about 1.3 bigger than the SM contribution. 
We show in Fig.~\ref{DMBDMFV.FIG} 
the region in MFV parameter space that satisfies this constraint, ignoring the gluino contribution. 

As pointed out in the previous subsection, in general we expect in the U(2) model,
$\tilde d_{R}\tilde s_{R}$ mixing to be near maximal, in which case the gluino contribution to 
\bdmix\ can be sizable. The gluino contribution can then be important to both 
$\Delta m_d$ and $a_{\psi K_s}$. Taking this into account, we can write 
$a_{\psi K_s} = \sin{(2\beta + 2 \theta_d)}$,  where $\theta_d$ is 
the new phase in $M_{12}(B_d)$~\cite{Grossman:1997dd}, and we 
have~\cite{Eidelman:2004wy,BRANCOCP.BOOK}
\bea
a_{\psi K_s} &=& {\rm Im}(\lambda_{\psi K}) \ ,  \nonumber \\
\lambda_{\psi K} \equiv -\frac{q}{p} \frac{\bar A(\bbarpsik)}{A(\bpsik)} \ &,& \qquad 
\frac{q}{p} \equiv \sqrt{\frac{ M_{12}^* - \frac{i}{2} \Gamma_{12}^* }{ M_{12} - \frac{i}{2} \Gamma_{12}}} \ ,
\eea
with $M_{12}$ and $\Gamma_{12}$ given in Eq.~(\ref{M12BQ.EQ}).
(The ``$-$'' sign in $\lambda_{\psi K}$ is because the final state is CP odd.)
In our case, $\Gamma_{12} \ll M_{12}$, so that  
\beq
a_{\psi K_s} \approx \sin{\left(\arg{(M_{12})}\right)} \ ,
\eeq
where ``arg'' denotes the argument of the complex quantity. 

\begin{figure}
\dofigsize{2.75in}{2.5in}{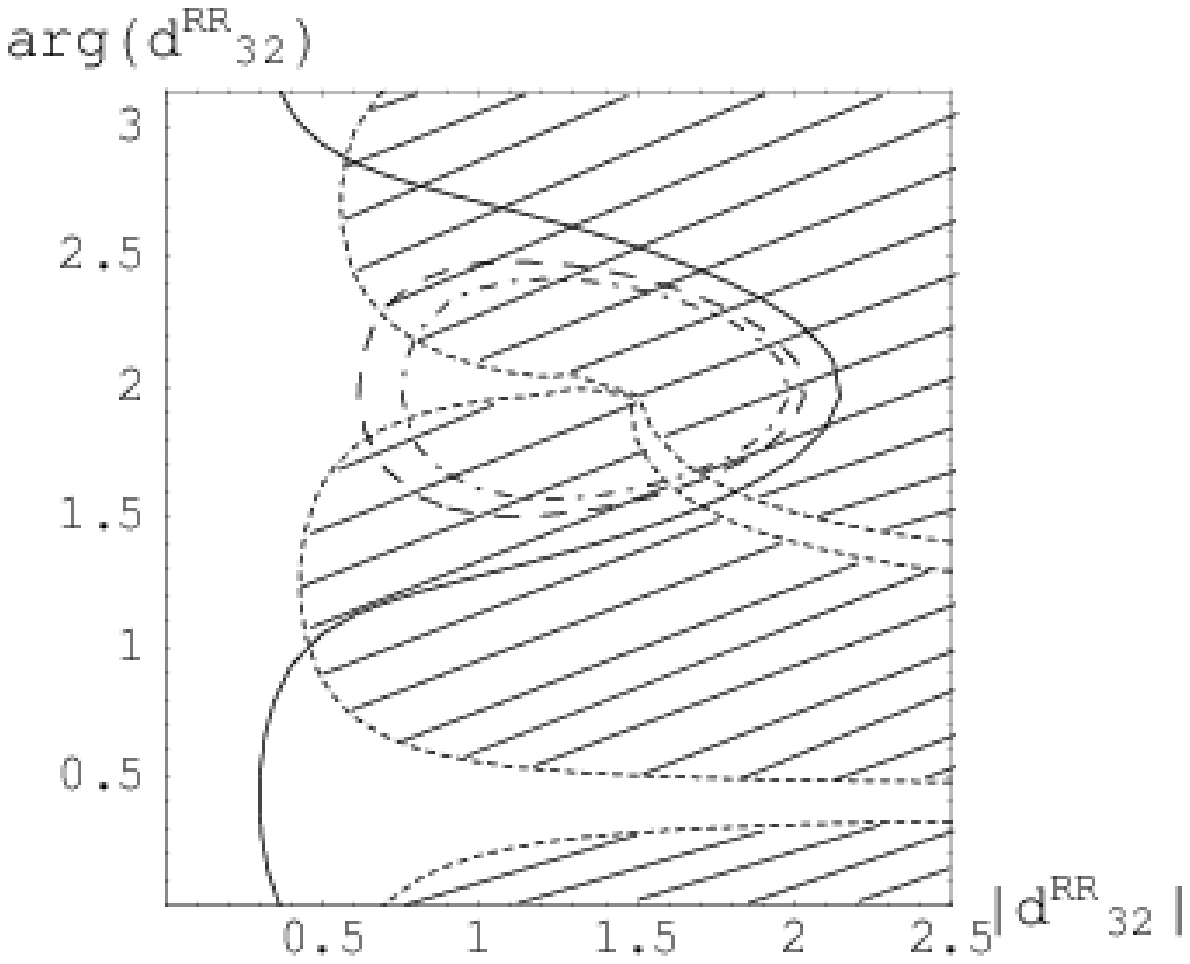}{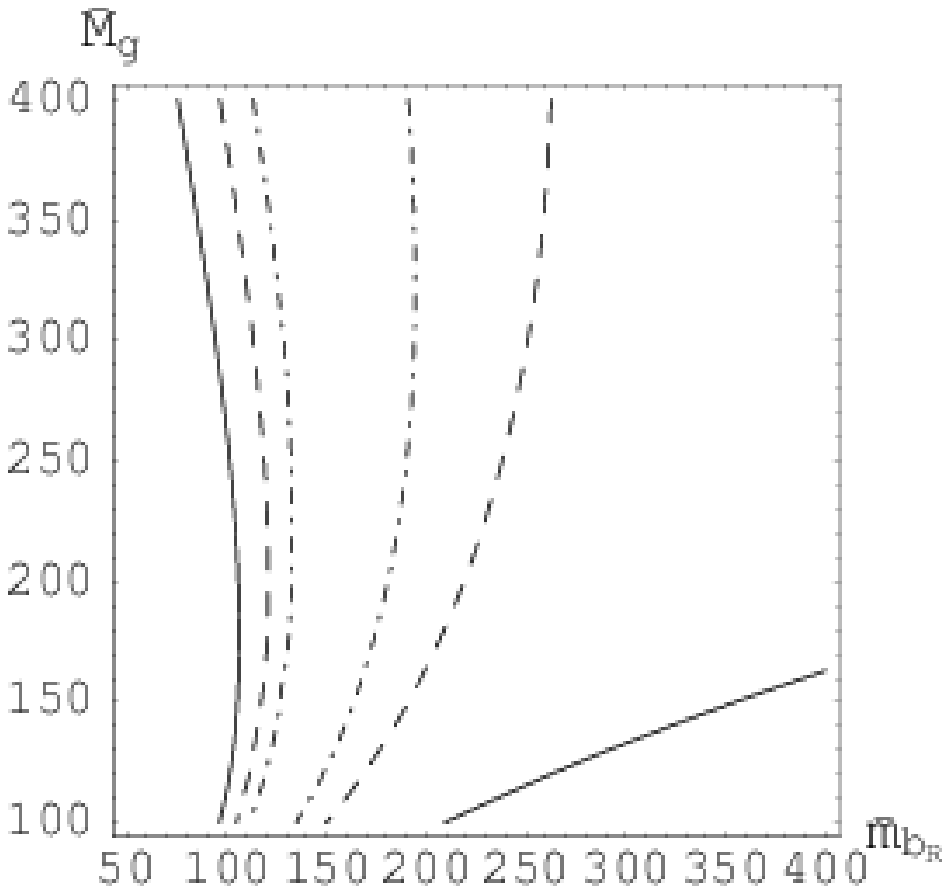}
\caption{For large $\tilde d_R\tilde s_R$ mixing, the (dash-dot,~dash,~solid) curves are (0.9,~1.0,~1.25) contours of $|(C_1+\tilde C_1)/C_1^{SM}|$, showing the NMFV contributions to \bdmix\ mixing relative to the SM. $d_{32}^{RR}$ is defined in Eq.~(\ref{DELDEF.EQ}).
The hatched region is excluded by $a_{\psi K_s}$.
Parameters not shown on a plot's axes are fixed as shown in Table~\ref{SPARAM.TAB}.
}
\label{M12BDNMFV.FIG}
\end{figure}

For the case when $\tilde d_R\tilde s_R$ mixing is large, we show the gluino contribution to \bdmix\ in 
Fig~\ref{M12BDNMFV.FIG}. 
The plot on the left also shows the constraint from $a_{\psi K_s}$, which is not shown in the plot on the right 
since almost the whole region shown is allowed. 
The region $(\pi < \arg{(\delta_{32}^{RR})} < 2\pi)$ is not shown since it is identical to the region $(0,~\pi)$.
From the figure, we see that in the large mixing case, the constraint on $\delta_{32}^{RR}$ is quite strong. 
However, if $\tilde d_R\tilde s_R$ mixing is small, the constraint on $\delta_{32}^{RR}$ from \bdmix\ mixing
is weak.

\subsection{\bsmix\ mixing}
\bsmix\ mixing has not yet been observed and the current experimental limit is 
$\Delta m_{B_s} > 14.4~ps^{-1}$ @ 95\% C.L.~\cite{Eidelman:2004wy}. The SM prediction is: 
$14~{\rm ps^{-1}} < \Delta m_{B_s} < 20~{\rm ps^{-1}}$~\cite{Anikeev:2001rk}. 
The SM prediction for the dilepton asymmetry $A_{ll}^{B_s}$ is small, around $10^{-4}$, 
cf. references in Ref.~\cite{Randall:1998te}.

\bsmix\ mixing depends quite sensitively on $\delta_{32}^{RR}$, and for the region 
in Fig.~\ref{M12BDNMFV.FIG} allowed by \bdmix\ mixing, we find 
$\Delta m_{B_s} \approx 22~{\rm ps^{-1}}$ and $A_{ll}^{B_s} \approx 5\times 10^{-4}$.
This $\Delta m_{B_s}$ is a little higher than the SM prediction, although may be within 
the SM allowed range, given uncertainties. 

\begin{figure}
\dofigs{3in}{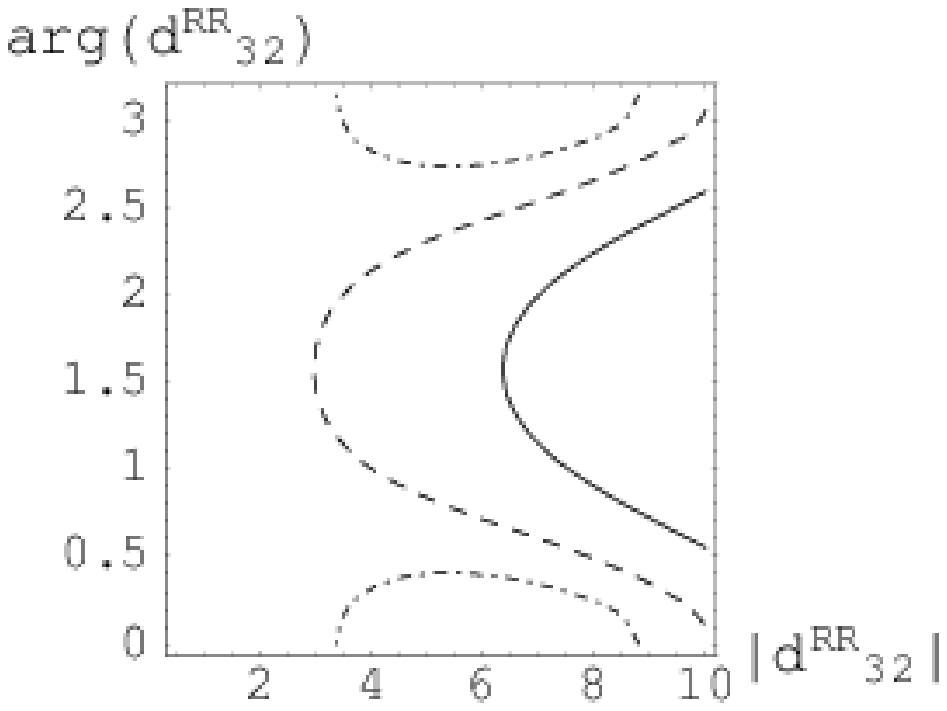}{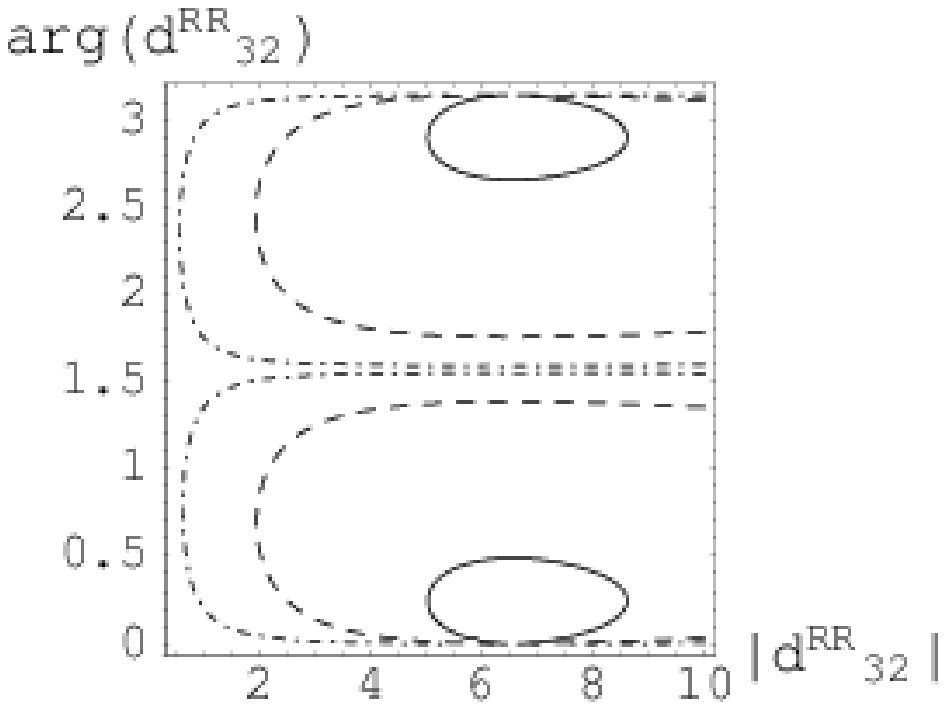}
\caption{For small $\tilde d_R\tilde s_R$ mixing, the (dash-dot,~dash,~solid) curves are 
(15,~25,~40~${\rm ps^{-1}}$) contours 
of $\Delta m_{B_s}$ (left), and ($10^{-4}$, $10^{-3}$ and $10^{-2}$) contours 
of $|A_{ll}^{B_s}|$ (right). $d_{32}^{RR}$ is defined in Eq.~(\ref{DELDEF.EQ}).
Parameters not shown on a plot's axes are fixed as shown in Table~\ref{SPARAM.TAB}.
}
\label{BSMIXRR.FIG}
\end{figure}

As we pointed out in the previous subsection, if $\tilde d_{R}\tilde s_{R}$ mixing is small,
then the \bdmix\ mixing constraints on $\delta_{32}^{RR}$ becomes weak. If such is the case,  
there are essentially no constraints from \bdmix\ mixing, and we show contours of 
$\Delta m_{B_s}$ and $A_{ll}^{B_s}$ in Fig.~\ref{BSMIXRR.FIG}. We show only the range
$(0< \arg{(\delta_{32}^{RR})} < \pi)$, since the $(\pi,~2\pi)$ range is identical to this.   
It can be seen that $\Delta m_{B_s}$ can increase significantly above the SM prediction. 
The projected Run II sensitivity for $\Delta m_{B_s}$ at the Tevatron with $2~{\rm fb^{-1}}$ is around 
$40~{\rm ps^{-1}}$~\cite{Anikeev:2001rk}, and can probe a significant region of U(2) parameter space. 
If a higher value of $\Delta m_{B_s}$ is measured than what the SM predicts, it would indicate the presence
of new physics. 
Measuring $A_{ll}^{B_s}$ can also significantly constrain $\delta_{32}^{RR}$ as can be seen from 
Fig.~\ref{BSMIXRR.FIG}~(right).

\section{$\Delta B=1$ FCNC processes}
\label{DB1.SEC}
\subsection{Effective Hamiltonian}
The $\Delta B=1$ effective Hamiltonian at a scale $\mu$ in the operator produce expansion~(OPE) 
is~\cite{Buchalla:1995vs,Buras:1998ra,Buras:1994dj}
\beq		
{\cal H}^{eff}_{\Delta B=1} = -\frac{G_F}{\sqrt{2}}V_{ts} V_{tb}^* \left(\sum_{i=1...6,9,10}\!\!\!\!\! C_i(\mu) O_i(\mu) + C_{7\gamma}(\mu) O_{7\gamma}(\mu) + C_{8g}(\mu) O_{8g}(\mu) \right) \ ,
\eeq
with
\bea
O_1    & = & (\bar{s}_{\alpha}  c_{\beta })_{V-A} (\bar{c}_{\beta }  b_{\alpha})_{V-A}, \nonumber \\
O_2    & = & (\bar{s} c)_{V-A}  (\bar{c} b)_{V-A},  \nonumber   \\
O_3    & = & (\bar{s} b)_{V-A}\sum_q(\bar{q}q)_{V-A}, \nonumber \\
O_4    & = & (\bar{s}_{\alpha}  b_{\beta })_{V-A} \sum_q (\bar{q}_{\beta} q_{\alpha})_{V-A}, \nonumber   \\
O_5    & = & (\bar{s} b)_{V-A}\sum_q(\bar{q}q)_{V+A}, \nonumber \\
O_6    & = & (\bar{s}_{\alpha}  b_{\beta })_{V-A} \sum_q  (\bar{q}_{\beta }  q_{\alpha})_{V+A}, \nonumber \\
O_{7\gamma}    & = & \frac{e}{8\pi^2} m_b \bar{s}_\alpha \sigma^{\mu\nu} (1+\gamma_5) b_\alpha F_{\mu\nu}, \nonumber \\
O_{8g}    & = & \frac{g_s}{8\pi^2} m_b \bar{s}_\alpha \sigma^{\mu\nu} (1+\gamma_5)T^a_{\alpha\beta} b_\beta G^a_{\mu\nu}, \nonumber \\
O_9    & = & (\bar{s} b)_{V-A}  (\bar{e}e)_V,  \nonumber \\
O_{10} & = & (\bar{s} b)_{V-A}  (\bar{e}e)_A,
\eea
\noindent where, the subscript $(V\pm A)$ means $\gamma_\mu (1\pm \gamma_5)$, and $F^{\mu \nu}$, $G^{\mu \nu}$ 
are the electromagnetic and color field strengths, respectively.

The Wilson coefficients can be computed at the scale $M_W$~(the $W$ boson mass), 
and then run down to the scale $m_b$ (the $b$ quark mass). 
Below, when no scale is specified for the coefficients, it is understood to be at $m_b$, i.e., 
$C_{i} \equiv C_{i}(m_b)$. The coefficients when run down from $M_W$ to $m_b$ mix under 
renormalization, so that~\cite{Buras:xp}~\footnote{Here, as a first step, we use the leading order 
result. The next to leading order result can be found in Ref.~\cite{Chetyrkin:1996vx}.} 
\bea
C_j &=& \sum_{i=1}^{8} k_{ji} \eta^{a_i}, \ \ \ \ (j=1,...,6), \nonumber \\
C_{7\gamma} &=& \eta^{\frac{16}{23}} C_{7\gamma}(M_W) + \frac{8}{3} \left( \eta^\frac{14}{23} - \eta^\frac{16}{23}\right) C_{8g}(M_W)  + \sum_{i=1}^{8} h_i \eta^{a_i} C_2(M_W),  \nonumber \\
C_{8g} &=& \eta^{\frac{14}{23}} C_{8g}(M_W) + \sum_{i=1}^{8} \bar{h}_i \eta^{a_i} C_2(M_W),
\label{EVOL.EQ}
\eea
\noindent where $\eta \equiv \frac{\alpha_s(M_W)}{\alpha_s(m_b)} \approx 0.56$ and $h_i$, $\bar{h}_i$, $a_i$ and 
$k_{ji}$ are given in Ref.~\cite{Buchalla:1995vs}. In addition, the evolution equation for $C_9$ is given in 
Ref.~\cite{Buras:1994dj}, and $C_{10}$ is not renormalized.

Separating out the new physics contribution to the renormalization group evolution, i.e., Eq.~(\ref{EVOL.EQ}), 
we get
\bea
C_2 &=& C_2^{SM}, \nonumber \\
C_{7\gamma} &=& C_{7\gamma}^{SM} + 0.67 C_{7\gamma}^{new}(M_W) + 0.09 C_{8g}^{new}(M_W), \nonumber \\
C_{8g} &=& C_{8g}^{SM} + 0.70 C_{8g}^{new}(M_W),
\label{WIL.EQ}
\eea
in which the superscript ``SM'' indicates the contribution from the SM, and ``new'' from new physics.

{\bf SM contribution}: The SM $W^\pm$ contribution to $C_{7\gamma}(M_W)$ and $C_{8g}(M_W)$ are given by~\cite{Grinstein:vj,Demir:2001yz}
\bea
C_2^{SM}(M_W) &=& 1, \\
C_{7\gamma,8g}^{SM}(M_W) &=& \frac{3}{2} F^{LL}_{7,8}\left(\frac{m_t^2}{M_W^2}\right), \nonumber \\
\label{C78SM.EQ}
\eea
where $F^{LL}_{7,8}(x)$ are given in Appendix~\ref{AppLoopFcn}. Using Eq.~(\ref{EVOL.EQ}) we can compute 
$C_2^{SM}$, $C_{7\gamma}^{SM}$ and $C_{8g}^{SM}$.

In the following, we will discuss, in order, the new physics contribution arising from the charged-Higgs boson 
($H^\pm$), charginos ($\tilde{\chi}^\pm$) and gluinos ($\tilde{g}$). \\
{\bf Charged Higgs ($H^\pm$) contribution}: The charged-Higgs contribution to \bsg\ 
is given by~\cite{Bertolini:1990if,Demir:2001yz,Barbieri:1993av}
\bea
C_{7\gamma,8g}^{H}(M_W) &=& \frac{1}{2} \cot^2\beta F^{LL}_{7,8}\left(\frac{m_t^2}{M_H^2}\right) + \tilde{F}^{LL}_{7,8}\left(\frac{m_t^2}{M_H^2}\right), 
\label{C78H.EQ}
\eea
where $F^{LL}_{7,8}(x)$ and $\tilde{F}^{LL}_{7,8}(x)$ are given in Appendix~\ref{AppLoopFcn}. \\
{\bf Chargino ($\tilde{\chi}^\pm$) contribution}: The chargino-stop contribution can be comparable to the 
SM contribution for a light stop and chargino. In the scenario that we are considering, the stop mixing angle 
is negligibly small and $m_{\tilde{t}_L} \approx m_{\tilde{t}_1} \sim \tilde{m}_0$ and 
$m_{\tilde{t}_R} \approx m_{\tilde{t}_2} \sim M_{W}$. We therefore run the $\tilde{t}_1$ contribution from 
$\tilde{m}_0$ down to $M_W$ and evaluate the $\tilde{t}_2$ contribution at $M_W$. The chargino-stop contribution 
is~\cite{Bertolini:1990if,Demir:2001yz,Barbieri:1993av}
\bea
C_{7\gamma,8g}^{\tilde\chi\tilde{t}_1}(\tilde{m}_0) = -\sum_{j=1}^{2}\left[|\Gamma_L^{1j}|^2 \frac{M_W^2}{m_{\tilde{t}_1}^2} F_{7,8}^{LL}\left(\frac{m_{\tilde{t}_1}^2}{M_{\tilde\chi_j}^2}\right) + \gamma_{RL}^{1j}\frac{M_W}{M_{\tilde\chi_j}} F_{7,8}^{RL}\left(\frac{m_{\tilde{t}_1}^2}{M_{\tilde\chi_j}^2}\right) \right], \nonumber \\
C_{7\gamma,8g}^{\tilde\chi\tilde{t}_2}(M_W) = -\sum_{j=1}^{2}\left[|\Gamma_L^{2j}|^2 \frac{M_W^2}{m_{\tilde{t}_2}^2} F_{7,8}^{LL}\left(\frac{m_{\tilde{t}_2}^2}{M_{\tilde\chi_j}^2}\right) + \gamma_{RL}^{2j}\frac{M_W}{M_{\tilde\chi_j}} F_{7,8}^{RL}\left(\frac{m_{\tilde{t}_2}^2}{M_{\tilde\chi_j}^2}\right) \right],
\label{C78CH.EQ}
\eea  
where the loop functions $F_{7,8}^{RL}$ are given in Appendix~\ref{AppLoopFcn}, and $\Gamma_L^{ij}$ and 
$\gamma_{RL}^{ij}$ contain the stop and chargino mixing matrices. Explicit expressions for $\Gamma_L^{ij}$, 
$\gamma_{RL}^{ij}$ and the renormalization group equations to evolve 
$C_{7\gamma,8g}^{\tilde\chi\tilde{t}_1}(\tilde{m}_0)$ down to $M_W$ are given in Ref.~\cite{Demir:2001yz}. \\
{\bf Gluino ($\tilde{g}$) contribution}: In our NMFV scenario, the gluino contributions can be sizable since 
they couple with strong interaction strength. Furthermore, because the sbottom mixing angle is negligibly small, 
$m_{\tilde{b}_L} \approx m_{\tilde{b}_1} \sim \tilde{m}_0$ and $m_{\tilde{b}_R} \approx m_{\tilde{b}_2} \sim M_{W}$. 
Keeping only the $\frac{M_{\tilde{g}}}{m_b}$ enhanced piece, the gluino contribution is~\cite{Bertolini:1990if}
\bea
C_{7\gamma}^{\tilde{g}}(M_W) = - \frac{4\pi\alpha_s\sqrt{2}}{G_F V_{ts}^* V_{tb}} \frac{M_{\tilde{g}}}{m_b} \cos{\theta^{RL}_{32}} \sin{\theta^{RL}_{32}} e^{-i\gamma^{RL}_{32}} \frac{1}{9} \left[ \frac{1}{m_{\tilde{b}_2}^2} F_4\left( \frac{M_{\tilde{g}}^2}{m_{\tilde{b}_2}^2}\right) - \frac{1}{m_{\tilde{b}_1}^2} F_4\left( \frac{M_{\tilde{g}}^2}{m_{\tilde{b}_1}^2}\right) \right] , \nonumber \\
C_{8g}^{\tilde{g}}(M_W) = \frac{4\pi\alpha_s\sqrt{2}}{G_F V_{ts}^* V_{tb}} \frac{M_{\tilde{g}}}{m_b} \cos{\theta^{RL}_{32}} \sin{\theta^{RL}_{32}} e^{-i\gamma^{RL}_{32}} \frac{1}{8} \left[ \frac{1}{m_{\tilde{b}_2}^2} F_{\tilde{g}}\left( \frac{M_{\tilde{g}}^2}{m_{\tilde{b}_2}^2}\right) - \frac{1}{m_{\tilde{b}_1}^2} F_{\tilde{g}}\left( \frac{M_{\tilde{g}}^2}{m_{\tilde{b}_1}^2}\right) \right] , \ 
\label{C78GL.EQ}
\eea
\noindent where the mixing angle $\theta^{RL}_{32}$ and phase $\gamma^{RL}_{32}$ are defined in 
Appendix~\ref{AppMixAng}, and $F_4$ and $F_{\tilde{g}}$ are defined in Appendix~\ref{AppLoopFcn}. In the above 
equation, we have neglected the effect of running the $\tilde{b}_1$ contribution from $\tilde{m}_0$ to $M_W$ 
as the $\tilde{b}_2$ contribution is dominant.

The dominant new physics contribution is given by adding Eqs.~(\ref{C78H.EQ}),~(\ref{C78CH.EQ})~and~(\ref{C78GL.EQ}), 
which yields 
\bea
C_{7\gamma,8g}^{new}(M_W) = C_{7\gamma,8g}^{H}(M_W) + C_{7\gamma,8g}^{\tilde\chi}(M_W) + C_{7\gamma,8g}^{\tilde{g}}(M_W).  
\label{BSGNEWPHY.EQ}
\eea

In what follows we will discuss in detail the new physics contribution predicted by the U(2) model to the rare 
decay processes \bsg, \bsglue, \bsll\ and \bphik.

\subsection{\bsg, \bsglue}
The dominant operators contributing to \bsg\ and \bsglue\ are $O_2$, $O_{7\gamma}$ and $O_{8g}$. 
The decay branching ratio B.R.(\bsg), at leading order, normalized to the semi-leptonic $B.R.(\bclnu)\approx 10.5 \% $, 
is given by~\cite{Buras:xp,mybsgwk,Kagan:1998bh}
\beq
\frac{\Gamma(\bsg)}{\Gamma(\bclnu)}\Biggr|_{(E_\gamma > (1-\delta)E_\gamma^{max})} = \frac{6\alpha}{\pi}\frac{1}{g(\frac{m_c}{m_b})} \left| \frac{V^*_{ts} V_{tb}}{V_{cb}} \right| ^2 |C_{7\gamma}|^2,
\label{BRBSG.EQ}
\eeq
\noindent where $g(z)\equiv 1-8z^2+8z^6-z^8-24z^4 \ln(z)$ is a phase space function, and $\delta$ is the 
fractional energy cut, i.e., only photon energy ${E_{\gamma} > (1 - \delta) E_{\gamma}^{max}}$ is accepted. 

The CP asymmetry in \bsg\ is given by~\cite{Kagan:1998bh}
\bea
A_{CP}^{\bsg}(\delta) &=&  \frac{\Gamma(\bar B_d \to X_s \gamma) - \Gamma(B_d \to X_{\bar s} \gamma)} {\Gamma(\bar B_d \to X_s \gamma) + \Gamma(B_d \to X_{\bar s} \gamma) }\Biggr|_{E_{\gamma} > (1 - \delta) E_{\gamma}^{max}} \ , \nonumber \\
&=& \frac{1}{|C_{7\gamma}|^2}\{ {a_{27}(\delta){\rm Im}[C_2C_{7\gamma}^*] + a_{87}(\delta){\rm Im}[C_{8g}C_{7\gamma}^*] + a_{28}(\delta){\rm Im}[C_2C_{8g}^*]} \}.
\label{ACPBSG.EQ}
\eea
\noindent For $\delta = 0.15$, which is a typical experimental cut, we use $a_{27} = 0.0124$, $a_{87} = -0.0952$ 
and $a_{28} = 0.0004$~\cite{Kagan:1998bh}. 

The experimentally measured~\cite{HFAG} branching ratio is 
$B.R.(\bsg) = (3.52^{+0.3}_{-0.28}) \times 10^{-4}$. 
In the SM, we have $C_2^{SM} \approx 1.11$, $C_{7\gamma}^{SM} \approx -0.31$ and 
$C_{8g}^{SM} \approx -0.15$. 
The SM prediction for B.R.(\bsg), which depends on $|C_{7\gamma}|$, cf. Eq.~(\ref{BRBSG.EQ}), is 
largely consistent with experiment, and new physics contributions to $|C_{7\gamma}|$ is
constrained by this branching ratio. 
In the context of SUSY this has been analyzed, for example, in 
Refs.~\cite{Bertolini:1990if,Buras:xp,mybsgwkbr}. 

The SM CP asymmetry in \bsg\ is of the order of 1\%, so that a larger CP asymmetry measured 
would imply new physics~\cite{Kagan:1998bh}. 
The present limit at 95~\% C.L. is~\cite{HFAG,Nishida:2003yw,Aubert:2004hq}
$-0.07 < A_{CP}^{\bsg} < 0.07$. 

The B.R.(\bsglue) is obtained simply from Eq.~(\ref{BRBSG.EQ})
\beq
\frac{\Gamma(\bsglue)}{\Gamma(\bclnu)} = C(R) \frac{6\alpha_s}{\pi}\frac{1}{g(\frac{m_c}{m_b})} \left| \frac{V^*_{ts} V_{tb}}{V_{cb}} \right| ^2 |C_{8g}|^2,
\label{BRBSGLUE.EQ}
\eeq
\noindent where the SU(3) quadratic Casimir $C(R)=4/3$.
The B.R.(\bsglue) has large experimental and theoretical uncertainties and Ref.~\cite{Kagan:1997qn} suggests 
that the data might prefer a B.R. value of around $10\%$.

\begin{figure}
\dofigs{2.5in}{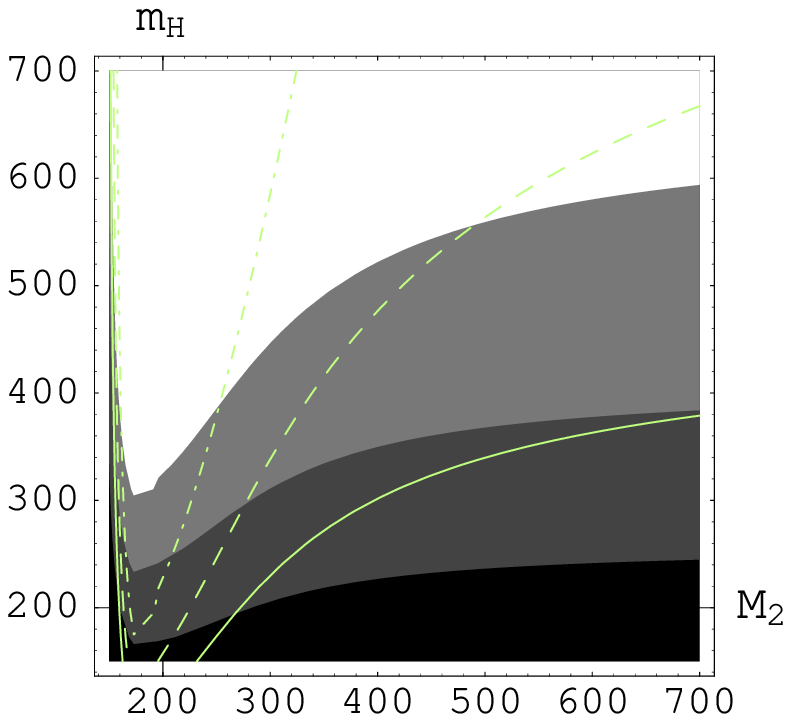}{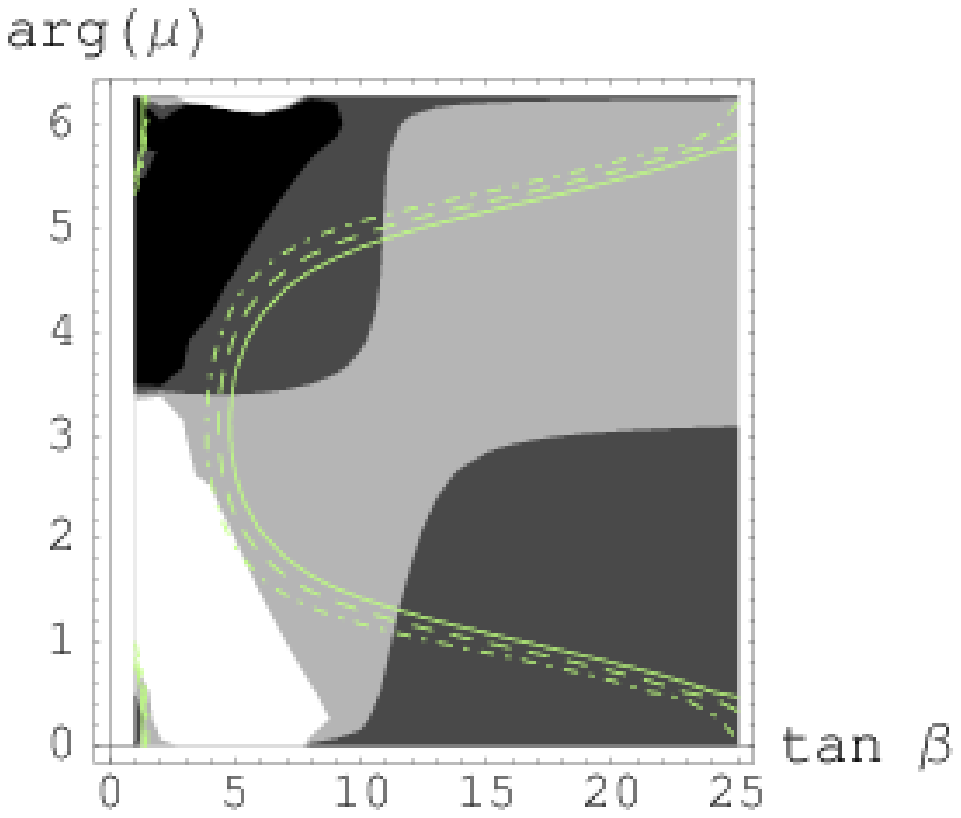}
\caption{The boundaries between the shaded regions show (2.5,3,3.5)\% (darkest to lightest) contours of $A_{CP}^{\bsg}$ 
as a function of $m_H$ and $M_2$ (left), and, (-3.5,0,3.5)\% (darkest to lightest) contours of $A_{CP}^{\bsg}$ as
a function of $\tan\beta$ and $\arg{(\mu)}$ (right).
Superimposed is the experimental $2\,\sigma$ allowed contours of B.R.(\bsg). 
Parameters not shown on a plot's axes are fixed as shown in Table~\ref{SPARAM.TAB}.}
\label{BSGACPMHM2.FIG}
\end{figure}

\begin{figure}
\dofigs{3.2in}{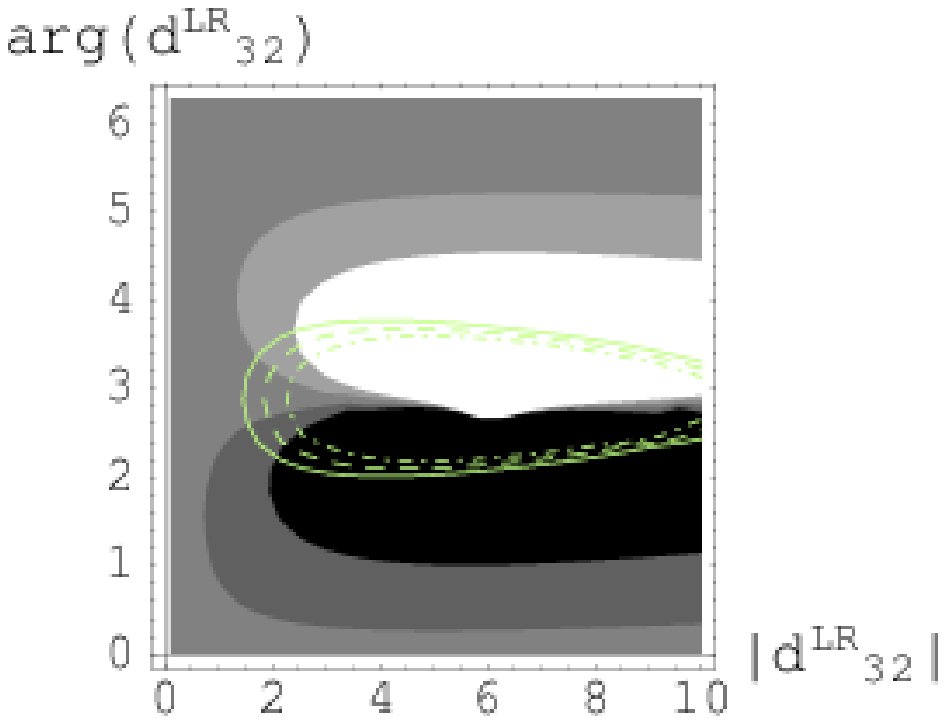}{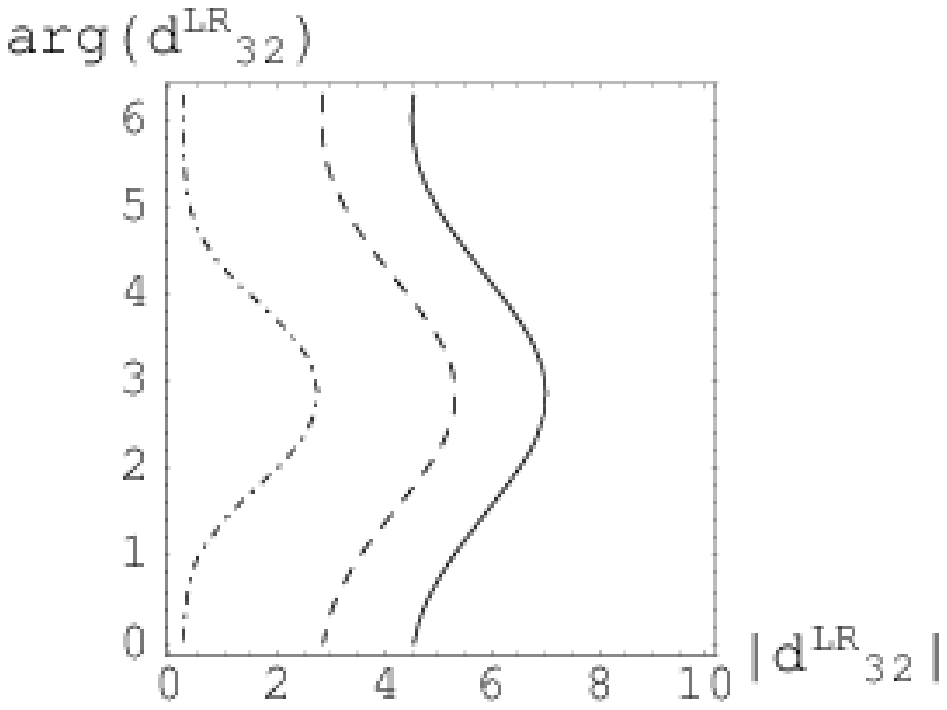}
\caption{The boundaries between the shaded regions show (-7,-3,3,7)\% (darkest to lightest) contours of $A_{CP}^{\bsg}$ (left) with experimentally allowed $2\,\sigma$ contours of B.R.(\bsg) superimposed, and, 1\,\%, 7.5\,\% and 15\,\% contours of predicted B.R.(\bsglue) (right).
Parameters not shown on a plot's axes are fixed as shown in Table~\ref{SPARAM.TAB}. }
\label{BSGACPLR.FIG}
\end{figure}

Figs.~\ref{BSGACPMHM2.FIG}~and~\ref{BSGACPLR.FIG} show the interplay between the 
$W^\pm$, $H^\pm$, $\tilde{\chi}^\pm$ and $\tilde{g}$ contributions to \bsg, where the sum of these contributions 
to the magnitude of $C_{7\gamma}$ is constrained by B.R.(\bsg). The experimental data on B.R.(\bsg) 
allows (at $2\,\sigma$) the region bounded by the contours shown in the figures. In the plots, the parameters
are as given in Table~\ref{SPARAM.TAB}, and some relevant ones are varied as shown in the figures.

To illustrate the dependence on the MFV parameters, we consider for example, in Fig.~\ref{BSGACPMHM2.FIG}, 
the dependence of $A_{CP}^{\bsg}$ and B.R.(\bsg) as a function of $m_H$ and $M_2$ (left) 
and as a function of $\tan\beta$ and $\arg{(\mu)}$ (right), 
for the choice of parameters shown in Table~\ref{SPARAM.TAB}.
The experimental $2\,\sigma$ allowed contours of B.R.(\bsg) are also shown.
In Fig.~\ref{BSGACPLR.FIG}~(left), the shaded regions show $A_{CP}^{\bsg}$ as a function of the magnitude 
and argument of $d_{32}^{RL}$, the dimensionless $\mathcal{O}(1)$ coefficient defined in 
Eq~(\ref{DELNUM.EQ}).
In Fig.~\ref{BSGACPLR.FIG}~(right), we show contours of B.R.(\bsglue), and a B.R. of up to about 15\% can
be accommodated in this model.

\subsection{\bsll}
The dominant operators contributing to \bsll\ ($\ell = e,\mu$) are 
$O_{7\gamma}$, $O_9$ and $O_{10}$. It is usual to define
\bea
C_9(\mu) &\equiv& \frac{\alpha}{2\pi} \tilde{C}_9(\mu), \nonumber \\
C_{10} &\equiv& \frac{\alpha}{2\pi} \tilde{C}_{10}, \nonumber
\eea
and
\beq
\hat{s} \equiv \frac{(p_{\ell^+}+p_{\ell^-})^2}{m_b^2}.
\eeq
The (differential) partial width $\frac{d}{d\hat s}\Gamma(\bsll)$, normalized to $\Gamma(\bclnu)$, is given 
by~\cite{Buras:1994dj}:
\bea
R(\hat{s})\equiv \frac{\frac{d}{d\hat s} \Gamma (\bsll)}{\Gamma (\bclnu)} = \frac{\alpha^2}{4\pi^2}
\left|\frac{V_{ts}^* V_{tb}}{V_{cb}}\right|^2 \frac{(1-\hat s)^2}{f(\frac{m_c}{m_b})\kappa(\frac{m_c}{m_b})} \left[(1+2\hat s)\left(|\tilde{C}_9^{eff}|^2 + |\tilde{C}_{10}|^2\right) + \right. \nonumber \\
\left. 4 \left( 1 + \frac{2}{\hat s}\right) |C_{7\gamma}|^2 + 12~{\rm Re}(C_{7\gamma} \tilde{C}_9^{eff}) \right],
\eea
\noindent where $f$ and $\kappa$ are phase space functions and $\tilde{C}_9^{eff}$ is the QCD 
corrected $\tilde C_9$, given in terms of $\tilde C_9$ and $C_i$~($i$=1...6)~\cite{Buras:1994dj}. 
Integrating this we get the prediction for the decay branching ratios and we show this in 
Table~\ref{BSLLDATA.TAB} for the SM along with the 
experimental result~\cite{HFAG}. We choose the lower limit on the integration to correspond to
a typical experimental choice, $(p_{\ell^+}+p_{\ell^-})^2 > (0.2~{\rm GeV})^2$. 
Since the rate of \bsll\ is down by the square of the electromagnetic coupling constant compared 
to \bsg, the experimental errors are comparatively larger. 
\begin{table}[h]
\centering
\begin{tabular}{||c|c|c||}
\hline
 & Experiment\cite{HFAG} & SM prediction \cr
\hline
$B.R.(\bsll)$ & $4.46^{+0.98}_{-0.96} \times 10^{-6}$ & $5.3 \times 10^{-6}$ \cr
\hline
\end{tabular}
\caption{The current data for \bsll.}
\label{BSLLDATA.TAB}
\end{table}

\begin{figure}
\dofig{3.5in}{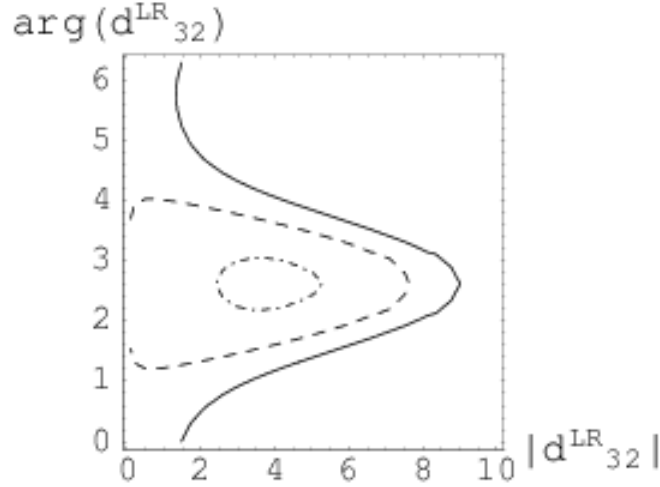}
\caption{The (dash-dot,~dash,~solid) curves are $(5.25,~6.25,~7.25)\times 10^{-6}$ contours of 
\bsll.
Parameters not shown on the plot's axes are fixed as shown in Table~\ref{SPARAM.TAB}.}
\label{BLLBRCONTLR.FIG}
\end{figure}
Fig.~\ref{BLLBRCONTLR.FIG} shows the contours of B.R.(\bsll) as a function of $d_{32}^{RL}$. 
Compared to B.R.(\bsg), cf. Fig.~\ref{BSGACPLR.FIG}~(left), the \bsll\ constraint is not very 
stringent right now, and improved statistics at the B-factories could place tighter constraints 
on the parameter space.

\subsection{\bphik}
The decay \bphik\ ($b\to ss\bar{s}$ at the quark level) can be a sensitive probe of new physics since 
the leading order SM contribution is one-loop suppressed, and loop processes involving heavy SUSY 
particles can contribute significantly. However, the computation of B.R.(\bphik) suffers 
from significant theoretical uncertainties in calculating the hadronic matrix elements. 
We follow the factorization approach, details of which are presented in 
Ref.~\cite{Beneke:2001ev}. 
The theoretical uncertainties largely cancel in the CP asymmetry, and is
therefore a good probe of new physics. 

The CP asymmetry in \bphik\ is defined by
\bea
A_{CP}^{\bphik} &\equiv& \frac{\Gamma\left(\bar{B}_d(t)\to \phi K_s\right) - \Gamma\left(B_d(t)\to \phi K_s \right)}{\Gamma\left(\bar{B}_d(t)\to \phi K_s\right) + \Gamma\left(B_d(t)\to \phi K_s \right)} \\
&=& -C_{\phi K}\cos{\left(\Delta m_{B_d}\,t\right)} + S_{\phi K}\sin{\left(\Delta m_{B_d}\,t\right)} \ ,
\eea
\noindent where 
\bea
C_{\phi K} &\equiv& \frac{1-|\lambda_{\phi K}|^2}{1+|\lambda_{\phi K}|^2} \ , \nonumber \\
S_{\phi K} &\equiv& \frac{2\,{\rm Im}{\left(\lambda_{\phi K}\right)}}{1+|\lambda_{\phi K}|^2} \ ,  \nonumber \\
\lambda_{\phi K} &\equiv& - e^{-2i(\beta + \theta_d)} \frac{\bar{A}(\bphik)}{A(\bbarphik)} \ , \nonumber
\eea
where $B_d(t)$ represents the state that is a $B_d$ at time $t=0$, $\Delta m_{B_d}$ is the \bdmix\  
mass difference, $\beta$ is the usual angle in the SM CKM unitarity triangle fits to the CP asymmetry in 
\bjpsik, and $\theta_d$ is any new physics contributions to \bdmix\ mixing ($\theta_d$ is discussed in 
Section~\ref{BDMIX.SEC}).   
The SM predicts that the CP asymmetry in \bphik\ and \bjpsik\ should be the same, i.e., $S_{\phi K} = \sin{2\beta}$.

The \bbarphik\ amplitude and partial decay width are given by~\cite{Kane:2002sp,Beneke:2001ev}: 
\bea
A(\bbarphik) &=& \sum_{p=u,c}\lambda_p\left[(a_3+a_4^p+a_5)-\frac{1}{2}(a_7+a_9+a_{10}^p) \right] \\
\Gamma(\bbarphik) &=& \frac{G_F^2 f_\phi^2 m_B^3}{32\pi}(F_1^{B\to K})^2 |A(\bbarphik)|^2 \left[\lambda(1,\frac{m_\phi^2}{m_B^2},\frac{m_K^2}{m_B^2})\right]^\frac{3}{2}
\eea
where the phase space function\footnote{We thank Liantao Wang for clarifying the expression for 
$\lambda(x,y,z)$.} $\lambda(x,y,z) \equiv x^2+y^2+z^2-2xy-2yz-2zx$, the $\phi$ decay constant 
$f_\phi = 237$~MeV, the form factor $F_1^{B\to K} = 0.38$ and $\lambda_p \equiv V_{pb}V_{ps}^*$. 
The SM $a_i$'s, in terms of the $C_i$'s, are given in Ref.~\cite{Beneke:2001ev} to which we
add the new physics contribution given in Eq.~(\ref{BSGNEWPHY.EQ}). We do not include the power-suppressed
weak annihilation operators and we refer the reader to Refs.~\cite{Beneke:2001ev}~and~\cite{Beneke:2003zv} 
for a more complete discussion. As explained in 
Section.~\ref{SUSYPARAM.SEC}, we are only including the $\delta_{32}^{RL}$ SUSY contribution, 
as this is the dominant one. The amplitude for the CP conjugate process \bphik\ is obtained 
by taking $\lambda_p \to \lambda_p^*$.

The current \bphik\ experimental average~\cite{HFAG} 
is summarized in Table.~\ref{BPHIKDATA.TAB}. The SM requires 
$S_{\phi K} = S_{J/\psi K} \equiv \sin{2\beta}$, but the experimental data has about a $2\,\sigma$ 
discrepancy between $S_{\phi K}$ and $S_{J/\psi K}$.\footnote{The significance of the discrepancy
between $S_{b\rightarrow s}$ and $S_{b\rightarrow c}$ is bigger, currently at about $3.5\,\sigma$, 
where $S_{b\rightarrow s}$ and $S_{b\rightarrow c}$ are the averages over all measured 
$b\rightarrow s$ (penguin) and $b\rightarrow c$ modes, respectively.}
Though not convincing yet, this could be an indication of 
new physics and we ask if this can be naturally explained in the theory we are considering.
\begin{table}[h]
\centering
\begin{tabular}{||c|c|c||}
\hline
 & Experiment~\cite{HFAG} & SM prediction \cr
\hline
$B.R.(\bphik)$ & $8.3^{+1.2}_{-1.0}\times 10^{-6}$ & $\sim 5\times 10^{-6}$ \cr
$S_{\phi K}$ & $0.34\pm 0.2$ & $0.725\pm 0.037$ \cr
$C_{\phi K}$ & $-0.04\pm 0.17$ & $0$ \cr
\hline
\end{tabular}
\caption{The current data for \bphik.}
\label{BPHIKDATA.TAB}
\end{table}

\begin{figure}
\dofigsthr{2in}{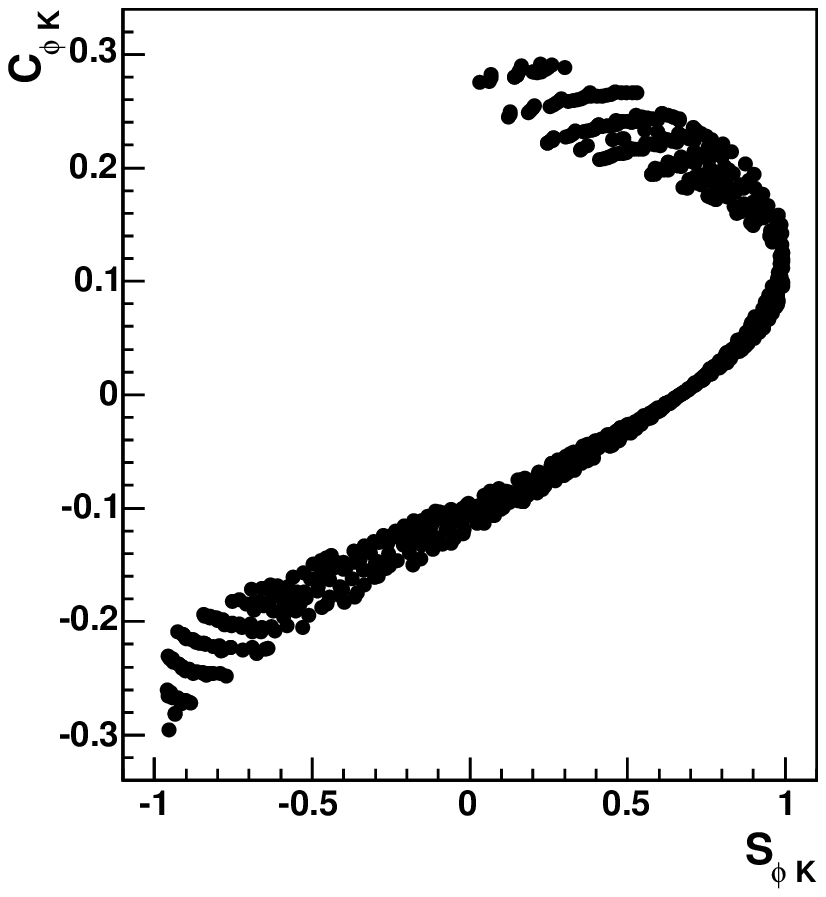}{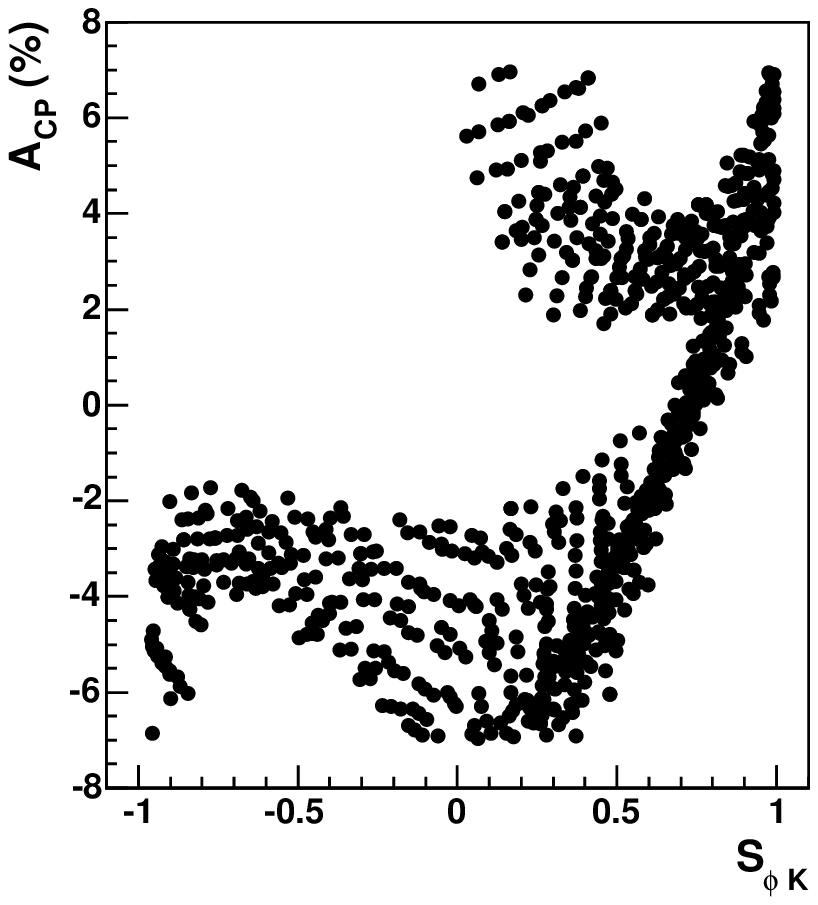}{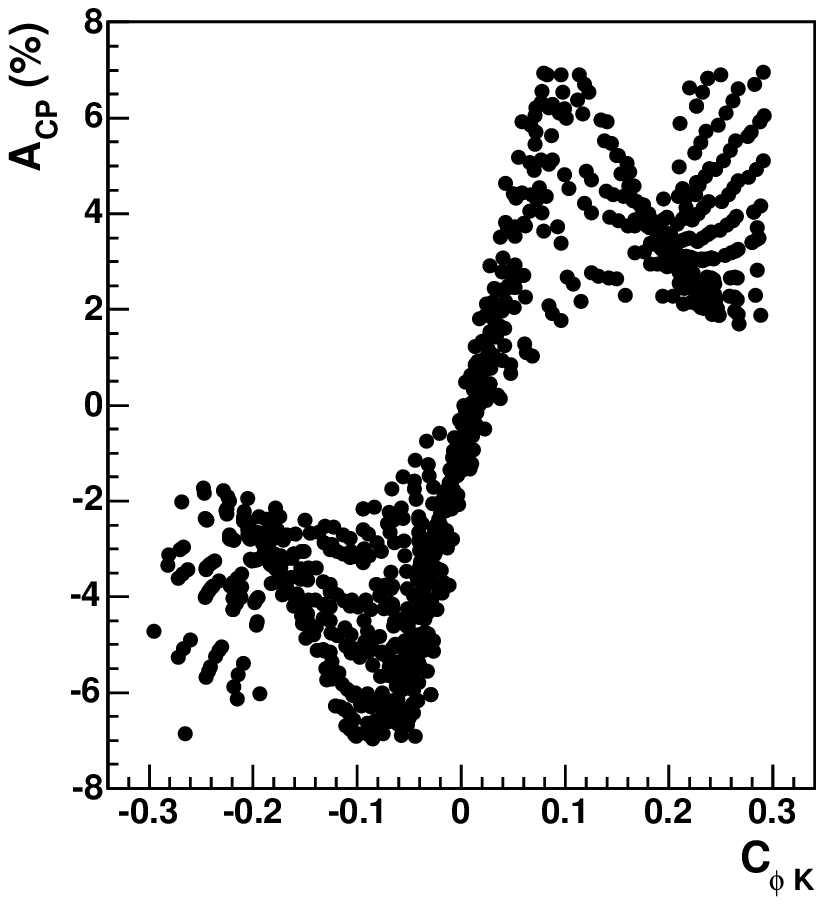}
\caption{$A_{CP}^{\bsg}$,  $S_{\phi K}$ and $C_{\phi K}$ for points
that satisfy all experimental constraints (within $2\,\sigma$), resulting from a scan
over $d_{32}^{LR}$ and ${\rm arg}(\mu)$, for small $\tilde d_R \tilde s_R$ mixing 
(negligible $\theta_d$).
All other parameters are fixed as shown in Table~\ref{SPARAM.TAB}.}
\label{BPHIKCPTH0.FIG}
\end{figure}

We showed in Section~\ref{BDMIX.SEC}, that if $\tilde d_R \tilde s_R$ mixing is small, there 
is no significant new phase in $M_{12}(B_d)$ (i.e., $\theta_d \approx 0$). 
For this case, we scan the parameter space $\arg(\mu)$, $|\delta_{32}^{RL}|$,
$\arg(\delta_{32}^{RL})$, and in Fig.~\ref{BPHIKCPTH0.FIG} show a scatter-plot of the 
points that satisfy all experimental constraints including B.R.(\bsg) and B.R.(\bphik). 
We find that it is possible to satisfy all experimental constraints including the recent
\bphik\ data shown in Table~\ref{BPHIKDATA.TAB} in the framework we are considering.
Furthermore, there are strong correlations between $A_{CP}^{\bsg}$, $S_{\phi K}$ and 
$C_{\phi K}$. As the accuracy of the experimental data improve, we can use these correlations
to (in)validate the choices that we make in our model.

\begin{figure}
\dofigsthr{2in}{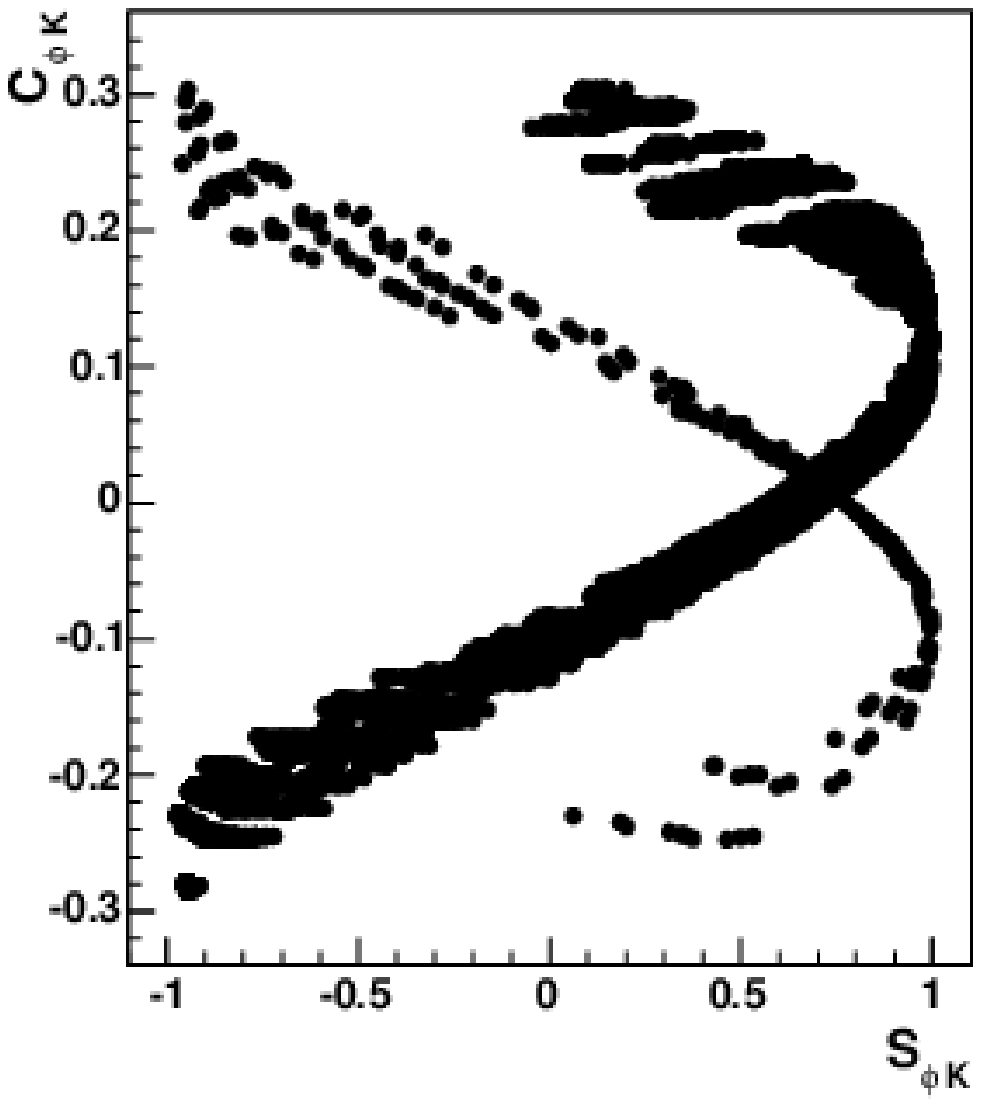}{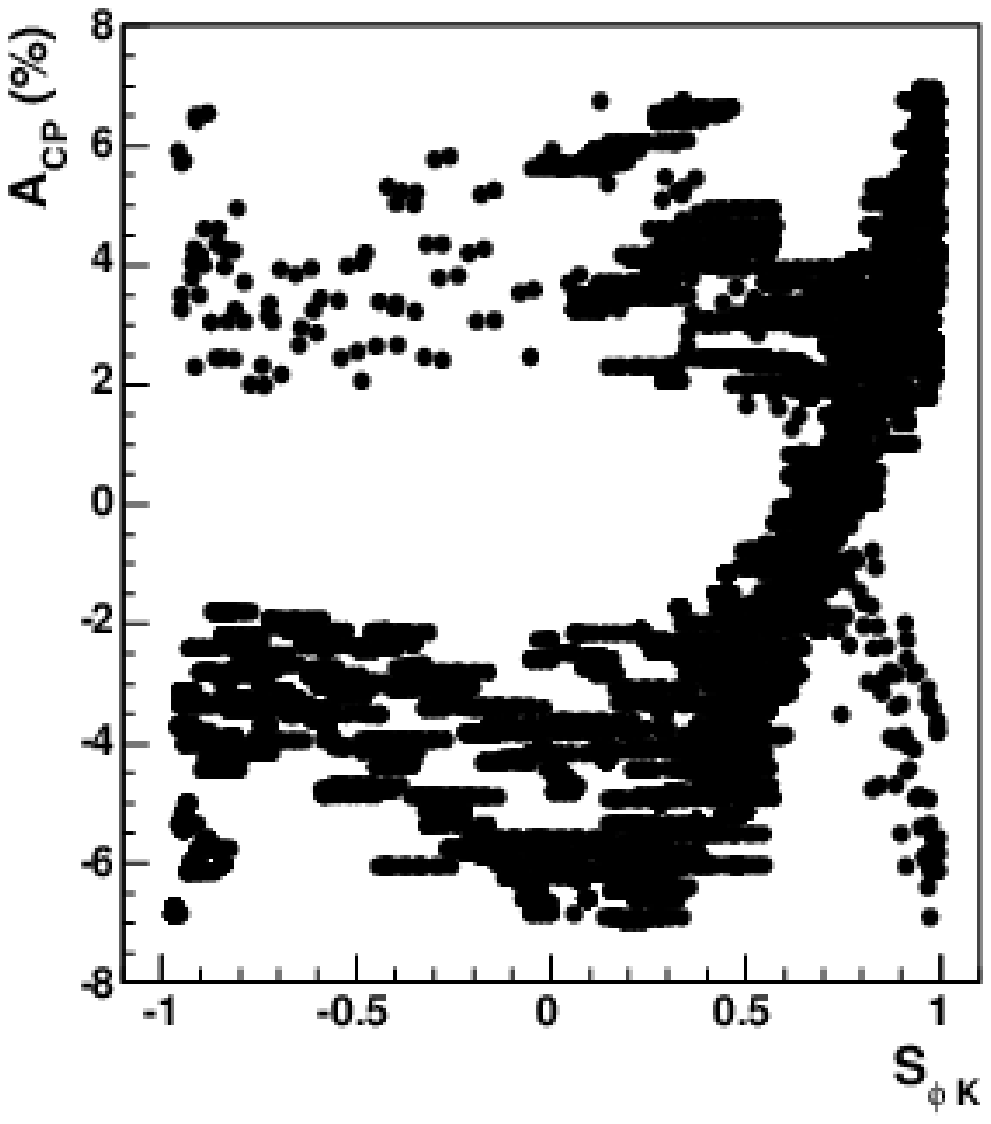}{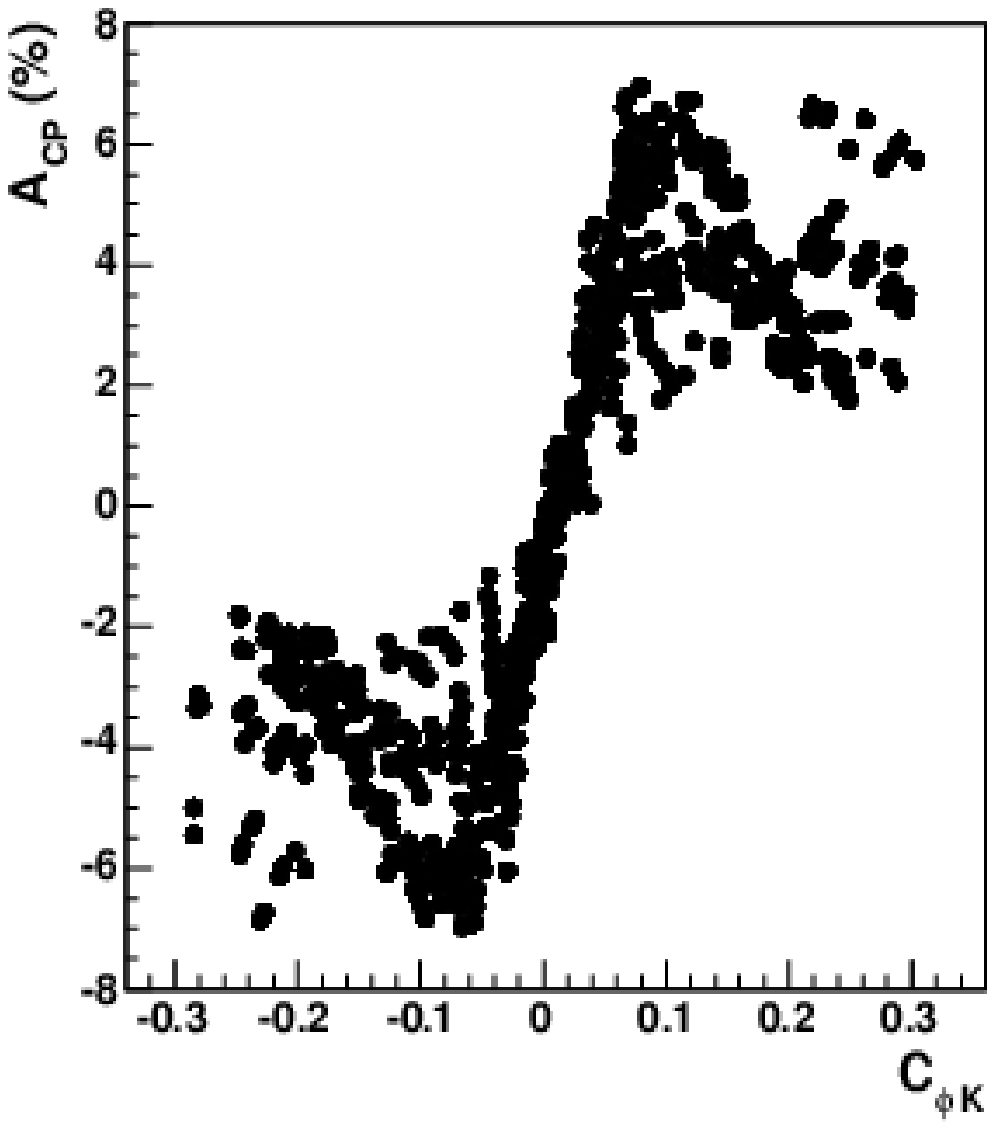}
\caption{$A_{CP}^{\bsg}$,  $S_{\phi K}$ and $C_{\phi K}$ for points
that satisfy all experimental constraints (within $2\,\sigma$), resulting from a scan
over $d_{32}^{LR,RR}$ and ${\rm arg}(\mu)$, for large $\tilde d_R \tilde s_R$ mixing 
(nonzero $\theta_d$).
All other parameters are fixed as shown in Table~\ref{SPARAM.TAB}.}
\label{BPHIKCP.FIG}
\end{figure}

Large $\tilde d_R \tilde s_R$ mixing can lead to a nonzero $\theta_d$ which depends on 
$|\delta_{32}^{RR}|$ as explained in Section~\ref{BDMIX.SEC}. 
We therefore include this new phase and perform a
scan over $|\delta_{32}^{RR}|$, $\arg(\delta_{32}^{RR})$, $\arg(\mu)$, $|\delta_{32}^{RL}|$ and 
$\arg(\delta_{32}^{RL})$.  
We show the points that satisfy all experimental constraints and the resulting
$A_{CP}^{\bsg}$, $S_{\phi K}$ and $C_{\phi K}$ in Fig.~\ref{BPHIKCP.FIG}.
We again see from Fig.~\ref{BPHIKCP.FIG} that $A_{CP}^{\bsg}$, $S_{\phi K}$ and 
$C_{\phi K}$ are strongly correlated, although the effect of the new phase in 
\bdmix\ mixing allows new regions of parameter space compared to the small mixing case 
shown in Fig.~\ref{BPHIKCPTH0.FIG}. Even in the case of
large mixing we find that it is possible to satisfy all experimental data including the
$S_{\phi K}$ and $C_{\phi K}$. One feature that we find in either large or small mixing
case is that sign~$(C_{\phi K})$ is positively correlated with sign~$(A_{CP}^{\bsg})$.
Thus, further data could shed light on the validity of the choices that we make in our
model.

\section{Conclusions}
\label{CONCL.SEC}
A supersymmetric U(2) theory has the potential to explain the gauge hierarchy and flavor 
problems in the SM. We assumed an effective SUSY mass spectrum just above the weak scale, 
the only relatively light scalars being the right handed stop and sbottom (weak scale masses).
We analyzed what such a hypothesis would imply for $K$ and $B$ meson observables by including
all the dominant contributions that can interfere in a certain observable.
Although for definiteness we considered a U(2) framework, our conclusions hold for any theory
with a similar SUSY mass spectrum and structure of the squark mass matrix. 
 
The CP violation parameter in Kaon mixing, $\epsilon_K$, can impose constraints on the 
MFV parameter space of our model, as we showed in Fig.~\ref{EPSKMFV.FIG}, while the 
gluino contribution to $\epsilon_K$ is negligible. 
There is sufficient room to accommodate the MFV contributions to $\epsilon_K$,
given the present uncertainty in the lattice computation of the Bag parameter $B_K$.

We find that \bdmix\ mixing and $a_{\psi K_s}$~($\sin{2\beta}$) can impose 
constraints on the supersymmetric U(2) theory.
In addition to the MFV contribution, if $\tilde d_R\tilde s_R$ mixing is large, 
the gluino contributions to \bdmix\ mixing can be significant leading to a 
strong constraint on the 32 entry of the RR squark mass matrix, $\delta_{32}^{RR}$, 
as shown in Fig.~\ref{M12BDNMFV.FIG}. Furthermore, in this case, there
is a new phase in the \bdmix\ mixing amplitude coming from the SUSY sector.
However, if $\tilde d_R\tilde s_R$ mixing is small, 
the constraint on $\delta_{32}^{RR}$ from \bdmix\ mixing is weak.

\bsmix\ mixing most sensitively depends on $\delta_{32}^{RR}$ in the SUSY U(2) theory. 
If $\delta_{32}^{RR}$ is unconstrained by \bdmix\ mixing (small $\tilde d_R\tilde s_R$ mixing), 
we showed that $\Delta m_{B_s}$ can be increased to quite large values 
(up to about $40~{\rm ps^{-1}}$), cf. Fig.~\ref{BSMIXRR.FIG} . 
The current and upcoming experiments can reach sensitivities required to see the SM prediction 
for $\Delta m_{B_s}$. Seeing a higher value, or not seeing a signal at all, might hint at some 
new physics of the type we are considering. We also presented expectations for the $B_s$ 
dilepton asymmetry, $A_{ll}^{B_s}$, which can constrain $\delta_{32}^{RR}$.

The experimental data on B.R.(\bsg) imposes a constraint on the SUSY theory. While 
satisfying this constraint, we showed that an enhancement in CP Violation in 
\bsg\ is possible. Since in the SM, $A_{CP}^{\bsg}$ is predicted to be less than $1\%$, 
if a much larger value is measured, it would clearly point to new physics. 
In Fig.~\ref{BSGACPLR.FIG} we presented the expectations for $A_{CP}^{\bsg}$ and 
B.R.(\bsglue) while varying the magnitude and phase of $\delta_{32}^{RL}$. 
We also presented expectations for B.R.(\bsll) in Fig.~\ref{BLLBRCONTLR.FIG}. 

The present experimental data on the CP violation in \bphik\ has about a $2\,\sigma$ deviation 
from the SM prediction, and it will be very interesting to see if this would persist with 
more data. We showed that such a deviation can be  accommodated in the framework we are 
considering, both for large or small $\tilde d_R\tilde s_R$ mixing. 
We showed, in Fig.~\ref{BPHIKCP.FIG}, that
$A_{CP}^{\bsg}$ can be enhanced significantly while satisfying all other experimental bounds 
including the present data on $S_{\phi K}$ and $C_{\phi K}$. 
In Figs.~\ref{BPHIKCPTH0.FIG}~and~\ref{BPHIKCP.FIG}, for small and large 
$\tilde d_R\tilde s_R$ mixing respectively, we see strong correlations between 
$A_{CP}^{\bsg}$, $S_{\phi K}$ and $C_{\phi K}$. Comparing these with upcoming data with 
improved precision could shed light on the validity of the choices that we make in our model. 

We conclude by remarking that the prospects are exciting for discovering SUSY in 
B-meson processes at current and upcoming colliders. Here, we showed this for a 
SUSY U(2) model. To unambiguously establish that it is a SUSY U(2) theory, and to 
determine the various SUSY breaking parameters, will require looking at a broad range of
observables. 

\section*{Acknowledgments}
We thank P.~Ko, U.~Nierste, K.~Tobe, J.~Wells and M.~Worah for many stimulating discussions, 
and, D.~Bortoletto and C.~Rott for discussions on the Tevatron bounds.
CPY thanks the hospitality of the National Center for Theoretical Sciences in Taiwan, ROC,
where part of this work was completed. 
SG acknowledges support from the high energy physics group at Northwestern 
University where this work was completed.
This work was supported in part by the NSF grant PHY-0244919.

%%%%%%%%%%%%%%%%%%%%%%%%%%%%%%%%%%%%%%%%%%%%%%%%%%%%%%%%%%%%%%%%%%%%%%%
\appendix
\section{Mixing Angles}
\label{AppMixAng}
The charged SU(2) Majorana gauginos $\tilde W_1$, $\tilde W_2$ can be combined to form the Dirac spinor
\beq
\tilde W^+ = \frac{1}{\sqrt{2}} \pmatrix{ {\tilde{W}^+}_\alpha \cr {\tilde{W}^{-\,\dot{\alpha}}}} \ ,
\eeq
where $\tilde W^\pm_\alpha = \tilde W_{1\alpha} \pm i \tilde W_{2\alpha}$.
The up and down type Higgsinos can be combined to form the Dirac spinor
\beq
\tilde H^+ = \pmatrix{ \tilde{H}_{u\,\alpha} \cr \tilde{H}_d^{\dot{\alpha}}} \ .
\eeq
The chargino mass terms can then be written as
\beq
{\cal L} \supset - \pmatrix{\overline{\tilde W^+} &  \overline{\tilde H^+}} \left( {\cal M}_\chi~P_L + {\cal M}_\chi^\dagger~P_R \right)   \pmatrix{\tilde W^+ \cr \tilde H^+ } \ ,
\eeq
where 
\beq
{\cal M}_\chi = \pmatrix{ M_2 &  \sqrt{2} \sin{\beta}~ m_W  \cr \sqrt{2} \cos{\beta}~ m_W & \mu } \ .
\eeq
We can go to the chargino mass eigen basis $(\chi_1 \ \chi_2)$  by making the rotations
\bea
P_{L,R} \pmatrix{\tilde W^+ \cr \tilde H^+} = \left({\cal C}_{L,R}\right)~ P_{L,R} \pmatrix{\tilde\chi_1^+ \cr \tilde\chi_2^+} \ , 
\eea
with the rotation matrices ${\cal C}_{L,R}$ given as
\beq
{\cal C}_\alpha = \pmatrix{\cos\theta_\alpha & - \sin\theta_\alpha ~ e^{- i \gamma_\alpha}  \cr \sin\theta_\alpha ~ e^{i \gamma_\alpha} & \cos\theta_\alpha } \pmatrix{e^{i\eta_\alpha} & 0 \cr 0 & e^{i\rho_\alpha}  } \ , 
\label{CDIAGCH.EQ}
\eeq
where the mixing angles and phases are~\cite{Demir:2001yz}
\bea
\gamma_L &=& -\arg{\left(M_2 + \mu \cot\beta\right)} \\
\tan{2\theta_L} &=& \frac{\sqrt{8}\, m_W \sin\beta |M_2 + \mu \cot\beta|}{M_2^2 + |\mu|^2+ 2\, m_W^2 \cos{2\beta} } \nonumber \\
\gamma_R &=& -\arg{\left(M_2 + \mu^* \tan\beta\right)} \nonumber \\
\tan{2\theta_R} &=& \frac{\sqrt{8}\, m_W \cos\beta |M_2 + \mu^* \tan\beta|}{M_2^2 - |\mu|^2 - 2\, m_W^2 \cos{2\beta} } \nonumber \\
\eta_R &=& \arg\left( c_R (M_2 c_L + \sqrt{2} m_W \sin\beta s_L e^{i\gamma_L} ) + s_R e^{-i\gamma_R} (\sqrt{2} m_W \cos\beta c_L + \mu s_L e^{i\gamma_L} ) \right) \nonumber \\
\rho_R &=& \arg\left( c_R (- \sqrt{2} m_W \cos\beta s_L e^{-i\gamma_L}-\mu c_L ) + s_R e^{i\gamma_R} ( -M_2 s_L e^{-i\gamma_L} + \sqrt{2} m_W \sin\beta c_L ) \right) \nonumber
\eea
with $0 \leq \theta_\alpha \leq \pi/2$ so that $M_{\alpha_1}^2 > M_{\alpha_2}^2$, and, $s_{L,R}\equiv \sin\theta_{L,R}$ 
and $c_{L,R}\equiv \cos\theta_{L,R}$. 

The sbottom mass terms are given as, cf. Eq.~(\ref{MSQ.EQ})
\beq
{\cal L} \supset -\pmatrix{\tilde b_L^* & \tilde b_R^*} \pmatrix{m_{3LL}^2+m_b^2+\Delta_L^d & (v_d A_b - \mu^* \tan{\beta}~m_b)^* \cr v_d A_b - \mu^* \tan{\beta}~m_b & m_{\tilde b_R}^2+m_b^2+\Delta_R^d} \pmatrix{\tilde b_L \cr \tilde b_R} \ ,
\eeq
where the $\Delta_{L,R}$ are the D-term contributions given as 
\beq
\Delta^d = \left(T_3 - Q_{EM}\sin^2\theta_W \right)\cos{2\beta} m_Z^2 \ .
\eeq
This is diagonalized by the rotation
\bea
\pmatrix{\tilde b_L \cr \tilde b_R} &=& \pmatrix{\cos{\theta_{\tilde b}} & -\sin{\theta_{\tilde b}} e^{-i\gamma_{\tilde b}} \cr\sin{\theta_{\tilde b}} e^{i\gamma_{\tilde b}} & \cos{\theta_{\tilde b}}} \pmatrix{\tilde b_1 \cr \tilde b_2} \nonumber \\
 &\equiv& \left( {\cal C}_{\tilde b} \right) \pmatrix{\tilde b_1 \cr \tilde b_2} \ ,
\label{CDIAGSB.EQ}
\eea
where the mixing angle and phase are given by
\bea
\tan{2\theta_{\tilde b}} &=& \frac{2 |v_d A_b - \mu^* \tan{\beta}~m_b|}{(m_{3LL}^2+\Delta_L^d) - (m_{\tilde b_R}^2+\Delta_R^d)} \ , \nonumber \\
\gamma_{\tilde b} &=& \arg{(v_d A_b - \mu^* \tan{\beta}~m_b)} \ .
\label{THDIAGSB.EQ}
\eea 
We have similar equations for stop mixing with obvious changes, in addition to the off 
diagonal term now being given as: ($v_u A_t - \mu^* \cot{\beta}~m_t$), and the stop mixing 
matrix denoted as ${\cal C}_{\tilde t}$. In our framework, owing to the smallness of the off 
diagonal RL mixing term compared to $m_{3LL}^2 \sim m_0^2$, we have small stop and sbottom mixing.
Furthermore, the sbottom mixing angle is negligibly small and we neglect its mixing effects. 
We thus have $\tilde b_1 \approx \tilde b_L$ and $\tilde b_2 \approx \tilde b_R$. 
The stop mixing angle, however, is not as small and so we include its effects.  

To compute the interaction vertices in the SuperKM basis, one could diagonalize the $6\times 6$ 
squark mass matrix. Since the off-diagonal entries in our case are small, we perform an approximate 
leading order diagonalization of the mass matrices shown in Eq.~(\ref{MSQ.EQ}). 

Focusing first on the $\tilde b_R\tilde s_L$ mixing,
\beq
{\cal L} \supset -\pmatrix{\tilde s_L^* & \tilde b_R^*} \pmatrix{m_{1LL}^2+\epsilon^2 m_{2LL}^2+m_s^2+\Delta_L^d & (v_d A_4^\prime \epsilon)^* \cr v_d A_4^\prime \epsilon & m_{\tilde b_R}^2+m_b^2+\Delta_R^d} \pmatrix{\tilde s_L \cr \tilde b_R} \ .
\eeq 
This is diagonalized by the rotation
\bea
\pmatrix{\tilde s_L \cr \tilde b_R} &=& \pmatrix{\cos{\theta_{32}^{RL}} & -\sin{\theta_{32}^{RL}} e^{-i\gamma_{32}^{RL}} \cr\sin{\theta_{32}^{RL}} e^{i\gamma_{32}^{RL}} & \cos{\theta_{32}^{RL}}} \pmatrix{\tilde q_2 \cr \tilde q_3} \nonumber \\
 &\equiv& \left( {\cal C}_{\tilde b_R \tilde s_L} \right) \pmatrix{\tilde q_2 \cr \tilde q_3} \ .
\eea
The mixing angle and phase are given by
\bea
\tan{2\theta_{32}^{RL}} &=& \frac{2 |v_d A \epsilon|}{(m_{1LL}^2+m_s^2+\Delta_L^d)-(m_{\tilde b_R}^2+m_b^2+\Delta_R^d)} \ , \nonumber \\
\gamma_{32}^{RL} &=& \arg{(v_d A \epsilon)} \ .
\eea

The $\tilde b_R\tilde s_R$ mixing is given similarly. We have the mass terms
\beq
{\cal L} \supset -\pmatrix{\tilde s_R^* & \tilde b_R^*} \pmatrix{m_{1RR}^2+m_s^2+\Delta_R^d & (\epsilon m_4^2)^* \cr \epsilon m_4^2 & m_{\tilde b_R}^2+m_b^2+\Delta_R^d} \pmatrix{\tilde s_R \cr \tilde b_R} \ ,
\eeq 
which is diagonalized by the rotation
\bea
\pmatrix{\tilde s_R \cr \tilde b_R} &=& \pmatrix{\cos{\theta_{32}^{RR}} & -\sin{\theta_{32}^{RR}} e^{-i\gamma_{32}^{RR}} \cr\sin{\theta_{32}^{RR}} e^{i\gamma_{32}^{RR}} & \cos{\theta_{32}^{RR}}} \pmatrix{\tilde q_2 \cr \tilde q_3} \nonumber \\
 &\equiv& \left( {\cal C}_{\tilde b_R \tilde s_R} \right) \pmatrix{\tilde q_2 \cr \tilde q_3} \ .
\label{ROTBRSR.EQ}
\eea
The mixing angle and phase are given by
\bea
\tan{2\theta_{32}^{RR}} &=& \frac{2 |\epsilon m_4^2|}{(m_{1RR}^2+m_s^2+\Delta_R^d)-(m_{\tilde b_R}^2+m_b^2+\Delta_R^d)} \ , \nonumber \\
\gamma_{32}^{RR} &=& \arg{(\epsilon m_4^2)} \ .
\label{ANGPHBRSR.EQ}
\eea

The $\tilde d_{L}\tilde s_{L}$ and $\tilde d_{R}\tilde s_{R}$ mixing terms are 
\beq
{\cal L} \supset -\pmatrix{\tilde d_{L,R}^* & \tilde s_{L,R}^*} 
\pmatrix{ m_1^2+m_d^2+\Delta_{L,R}^d  & i \epsilon^\prime m_5^2 \cr 
-i \epsilon^\prime m_5^2 &  m_1^2+m_s^2+\epsilon^2 m_2^2+\Delta_{L,R}^d}_{LL,RR} \pmatrix{\tilde d_{L,R} \cr \tilde s_{L,R}} \ .
\label{MDSLLRR.EQ}
\eeq
The matrices that diagonalizes these, ${\mathcal C}_{\tilde d_{L} \tilde s_{L}}$ and 
${\mathcal C}_{\tilde d_{R} \tilde s_{R}}$  are given analogous to 
Eq.~(\ref{ROTBRSR.EQ}), the angle and phase ($\theta_{12}^{LL}$, $\gamma_{12}^{LL}$) and
($\theta_{12}^{RR}$, $\gamma_{12}^{RR}$) given 
analogous to Eq.~(\ref{ANGPHBRSR.EQ}), and we will not write them down explicitly.
The diagonal entries are split only by $\mathcal{O}(\epsilon^2)$, and therefore this mixing is maximal 
in general. However, if $\epsilon^\prime m_5^2 \ll \epsilon^2 m_2^2$, this mixing can be small.

\section{Loop functions}
\label{AppLoopFcn}
The \bsg\ loop functions are given by
\bea
F^{LL}_7(x) &=& \left(\frac{x(7-5x-8x^2)}{36(x-1)^3} + \frac{x^2(3x-2)}{6(x-1)^4} \ln{x}\right) \ , \nonumber \\
F^{LL}_8(x) &=& \left(\frac{x(2+5x-x^2)}{12(x-1)^3} - \frac{3x^2}{6(x-1)^4} \ln{x}\right) \ , \nonumber \\
F^{RL}_7(x) &=& \left(\frac{5-7x}{6(x-1)^2} + \frac{x(3x-2)}{3(x-1)^3} \ln{x}\right) \ , \nonumber \\
F^{RL}_8(x) &=& \left(\frac{1+x}{2(x-1)^2} - \frac{x}{(x-1)^3} \ln{x}\right) \ , \nonumber \\
\tilde{F}^{LL}_7(x) &=& \left(\frac{x(3-5x)}{12(x-1)^2} + \frac{x(3x-2)}{6(x-1)^3} \ln{x}\right) \ , \nonumber \\
\tilde{F}^{LL}_8(x) &=& \left(\frac{x(3-x)}{4(x-1)^2} - \frac{x}{2(x-1)^3} \ln{x}\right) \ .
\eea
\bea
F_4(x) &=& \frac{x^2-1-2x\ln{x}}{2(x-1)^3} \nonumber \\
F_{\tilde{g}}(x) &=& -\frac{5x^2-18x+13+(9-x)\ln{x}}{3(x-1)^3}
\eea

The SM box functions are given by
\bea
S_0(x) &=& \frac{4x-11x^2+x^3}{4(1-x)^2} - \frac{3x^3\ln{x}}{2(1-x)^3} \ , \nonumber \\
S_0(x_1,x_2) &=& \frac{x_1 x_2}{4} \biggl{\{} \frac{x_1^2-8x_1+4}{(x_1-x_2)(x_1-1)^2} \ln{x_1} + 
\frac{x_2^2-8x_2+4}{(x_2-x_1)(x_2-1)^2} \ln{x_2} \nonumber \\
&&~~~\biggl. - \frac{3}{(x_1-1)(x_2-1)} \biggl{\}} \ .
\label{BOXSMS0.EQ}
\eea
The charged-Higgs and chargino box functions are given by
\bea
Y_1(r_\alpha,r_\beta,s_i,s_j) = \frac{r_\alpha^2}{(r_\beta-r_\alpha)(s_i-r_\alpha)(s_j-r_\alpha)}\ln{r_\alpha} + \frac{r_\beta^2}{(r_\alpha-r_\beta)(s_i-r_\beta)(s_j-r_\beta)}\ln{r_\beta} \nonumber \\
  + \frac{s_i^2}{(r_\alpha-s_i)(r_\beta-s_i)(s_j-s_i)}\ln{s_i} + \frac{s_j^2}{(r_\alpha-s_j)(r_\beta-s_j)(s_i-s_j)}\ln{s_j} \ , \nonumber \\
Y_2(r_\alpha,r_\beta,s_i,s_j) = \sqrt{s_i s_j} \left[ \frac{r_\alpha}{(r_\beta-r_\alpha)(s_i-r_\alpha)(s_j-r_\alpha)}\ln{r_\alpha} + \frac{r_\beta}{(r_\alpha-r_\beta)(s_i-r_\beta)(s_j-r_\beta)}\ln{r_\beta} \right. \nonumber \\
\left.  + \frac{s_i}{(r_\alpha-s_i)(r_\beta-s_i)(s_j-s_i)}\ln{s_i} + \frac{s_j}{(r_\alpha-s_j)(r_\beta-s_j)(s_i-s_j)}\ln{s_j} \right] \ .
\label{BOXHY.EQ}
\eea
from which various limiting cases can be obtained.

The gluino box integrals are given as
\bea
I_4 \equiv I_4(M_{\tilde{g}}^2,M_{\tilde{g}}^2,m_{\tilde{b}_R}^2,m_{\tilde{b}_R}^2) = \int \!\frac{d^4p}{(2\pi)^4}~\frac{1}{(p^2-M_{\tilde{g}}^2)(p^2-M_{\tilde{g}}^2)(p^2-m_{\tilde{b}_R}^2)(p^2-m_{\tilde{b}_R}^2)} \ , \nonumber \\
\tilde I_4 \equiv \tilde I_4(M_{\tilde{g}}^2,M_{\tilde{g}}^2,m_{\tilde{b}_R}^2,m_{\tilde{b}_R}^2) = \int \!\frac{d^4p}{(2\pi)^4}~\frac{p^2}{(p^2-M_{\tilde{g}}^2)(p^2-M_{\tilde{g}}^2)(p^2-m_{\tilde{b}_R}^2)(p^2-m_{\tilde{b}_R}^2)} \ ,
\label{BOXI4.EQ}
\eea
which we evaluate numerically using LoopTools~\cite{Hahn:1998yk} in Mathematica.

%%%%%%%%%%%%%%%%%%%%%%%%%%%%%%%%%%%%%%%%%%%%%%%%%%%%%%%%%%%%%%%%%%%%%%%%


\begin{thebibliography}{99}

%\cite{Cohen:1996vb}
\bibitem{Cohen:1996vb}
A.~G.~Cohen, D.~B.~Kaplan and A.~E.~Nelson,
``The more minimal supersymmetric standard model,''
Phys.\ Lett.\ B {\bf 388}, 588 (1996)
[hep-ph/9607394].
%%CITATION = HEP-PH 9607394;%%

%\cite{Pomarol:1995xc}
\bibitem{Pomarol:1995xc}
A.~Pomarol and D.~Tommasini,
``Horizontal symmetries for the supersymmetric flavor problem,''
Nucl.\ Phys.\ B {\bf 466}, 3 (1996)
[hep-ph/9507462].
%%CITATION = HEP-PH 9507462;%%

%\cite{Barbieri:1995uv}
\bibitem{Barbieri:1995uv}
R.~Barbieri, G.~R.~Dvali and L.~J.~Hall,
``Predictions From A U(2) Flavour Symmetry In Supersymmetric Theories,''
Phys.\ Lett.\ B {\bf 377}, 76 (1996)
[hep-ph/9512388];
%%CITATION = HEP-PH 9512388;%%
%\cite{Barbieri:1997tu}
%\bibitem{Barbieri:1997tu}
R.~Barbieri, L.~J.~Hall and A.~Romanino,
``Consequences of a U(2) flavour symmetry,''
Phys.\ Lett.\ B {\bf 401}, 47 (1997)
[hep-ph/9702315].
%%CITATION = HEP-PH 9702315;%%

\bibitem{mylgtanb}
%\cite{Degrassi:2000qf}
%\bibitem{Degrassi:2000qf}
G.~Degrassi, P.~Gambino and G.~F.~Giudice,
``B $\to$ X/s gamma in supersymmetry: Large contributions beyond the  leading order,''
JHEP {\bf 0012}, 009 (2000)
[hep-ph/0009337];
%%CITATION = HEP-PH 0009337;%%
%\cite{Carena:2000uj}
%\bibitem{Carena:2000uj}
M.~Carena, D.~Garcia, U.~Nierste and C.~E.~Wagner,
``b $\to$ s gamma and supersymmetry with large tan(beta),''
Phys.\ Lett.\ B {\bf 499}, 141 (2001)
[hep-ph/0010003].
%%CITATION = HEP-PH 0010003;%%

%\cite{Bertolini:1990if}
\bibitem{Bertolini:1990if}
S.~Bertolini, F.~Borzumati, A.~Masiero and G.~Ridolfi,
``Effects Of Supergravity Induced Electroweak Breaking On Rare B Decays And Mixings,''
Nucl.\ Phys.\ B {\bf 353}, 591 (1991).
%%CITATION = NUPHA,B353,591;%%

%\cite{Hagelin:1992tc}
\bibitem{Hagelin:1992tc}
J.~S.~Hagelin, S.~Kelley and T.~Tanaka,
``Supersymmetric flavor changing neutral currents: Exact amplitudes and phenomenological analysis,''
Nucl.\ Phys.\ B {\bf 415}, 293 (1994).
%%CITATION = NUPHA,B415,293;%%

\bibitem{mygenwk}
%\cite{Hall:1985dx}
%\bibitem{Hall:1985dx}
L.~J.~Hall, V.~A.~Kostelecky and S.~Raby,
``New Flavor Violations In Supergravity Models,''
Nucl.\ Phys.\ B {\bf 267}, 415 (1986);
%%CITATION = NUPHA,B267,415;%%
%\cite{Gabbiani:rb}
%\bibitem{Gabbiani:rb}
F.~Gabbiani and A.~Masiero,
``FCNC In Generalized Supersymmetric Theories,''
Nucl.\ Phys.\ B {\bf 322}, 235 (1989);
%%CITATION = NUPHA,B322,235;%%
%\cite{Gabbiani:1996hi}
%\bibitem{Gabbiani:1996hi}
F.~Gabbiani, E.~Gabrielli, A.~Masiero and L.~Silvestrini,
``A complete analysis of FCNC and CP constraints in general SUSY extensions of the standard model,''
Nucl.\ Phys.\ B {\bf 477}, 321 (1996)
[hep-ph/9604387];
%%CITATION = HEP-PH 9610323;%%
%\cite{Misiak:1997ei}
%\bibitem{Misiak:1997ei}
M.~Misiak, S.~Pokorski and J.~Rosiek,
``Supersymmetry and FCNC effects,''
Adv.\ Ser.\ Direct.\ High Energy Phys.\  {\bf 15}, 795 (1998)
[hep-ph/9703442].
%%CITATION = HEP-PH 9703442;%%

%\cite{Kane:2002sp}
\bibitem{Kane:2002sp}
G.~L.~Kane, P.~Ko, H.~b.~Wang, C.~Kolda, J.~h.~Park and L.~T.~Wang,
``B/d $\to$ Phi K(S) and supersymmetry,''
Phys.\ Rev.\ D {\bf 70}, 035015 (2004)
[hep-ph/0212092];
%%CITATION = HEP-PH 0212092;%%
%\cite{Agashe:2003rj}
%\bibitem{Agashe:2003rj}
K.~Agashe and C.~D.~Carone,
``Supersymmetric flavor models and the B $\to$ Phi K(S) anomaly,''
Phys.\ Rev.\ D {\bf 68}, 035017 (2003)
[hep-ph/0304229].
%%CITATION = HEP-PH 0304229;%%

%\cite{Affolder:1999wp}
\bibitem{Affolder:1999wp}
T.~Affolder {\it et al.}  [CDF Collaboration],
``Search for scalar top and scalar bottom quarks in p anti-p collisions  at
%s**(1/2) = 1.8-TeV,''
Phys.\ Rev.\ Lett.\  {\bf 84}, 5704 (2000)
[hep-ex/9910049];
%%CITATION = HEP-EX 9910049;%%
%\cite{Rott:2004is}
%\bibitem{Rott:2004is}
C.~Rott,
``Searches for the Supersymmetric Partner of the Bottom Quark,''
hep-ex/0410007.
%%CITATION = HEP-EX 0410007;%%

%\cite{Eidelman:2004wy}
\bibitem{Eidelman:2004wy}
S.~Eidelman {\it et al.}  [Particle Data Group Collaboration],
``Review of particle physics,''
Phys.\ Lett.\ B {\bf 592}, 1 (2004).
%%CITATION = PHLTA,B592,1;%%

%\cite{Buchalla:1995vs}
\bibitem{Buchalla:1995vs}
G.~Buchalla, A.~J.~Buras and M.~E.~Lautenbacher,
``Weak Decays Beyond Leading Logarithms,''
Rev.\ Mod.\ Phys.\  {\bf 68}, 1125 (1996)
[hep-ph/9512380].
%%CITATION = HEP-PH 9512380;%%

%\cite{Branco:1994eb}
\bibitem{Branco:1994eb}
G.~C.~Branco, G.~C.~Cho, Y.~Kizukuri and N.~Oshimo,
``Supersymmetric contributions to B0 - anti-B0 and K0 - anti-K0 mixings,''
Phys.\ Lett.\ B {\bf 337}, 316 (1994)
[hep-ph/9408229];
%%CITATION = HEP-PH 9408229;%%
%\cite{Branco:1995cj}
%\bibitem{Branco:1995cj}
%G.~C.~Branco, G.~C.~Cho, Y.~Kizukuri and N.~Oshimo,
``Searching for signatures of supersymmetry at B factories,''
Nucl.\ Phys.\ B {\bf 449}, 483 (1995).
%%CITATION = NUPHA,B449,483;%%

%\cite{Becirevic:2001jj}
\bibitem{Becirevic:2001jj}
D.~Becirevic {\it et al.},
``B/d anti-B/d mixing and the B/d $\to$ J/psi K(S) asymmetry in general  SUSY models,''
Nucl.\ Phys.\ B {\bf 634}, 105 (2002)
[hep-ph/0112303].
%%CITATION = HEP-PH 0112303;%%

%\cite{Buras:1997fb}
\bibitem{Buras:1997fb}
A.~J.~Buras and R.~Fleischer,
``Quark mixing, CP violation and rare decays after the top quark  discovery,''
Adv.\ Ser.\ Direct.\ High Energy Phys.\  {\bf 15}, 65 (1998)
[hep-ph/9704376].
%%CITATION = HEP-PH 9704376;%%

%\cite{Ko:2002ee}
\bibitem{Ko:2002ee}
P.~Ko, J.~h.~Park and G.~Kramer,
``B0 - anti-B0 mixing, B $\to$ J/psi K(S) and B $\to$ X/d gamma in general  MSSM,''
Eur.\ Phys.\ J.\ C {\bf 25}, 615 (2002)
[hep-ph/0206297].
%%CITATION = HEP-PH 0206297;%%

%\cite{Randall:1998te}
\bibitem{Randall:1998te}
L.~Randall and S.~f.~Su,
``CP violating lepton asymmetries from B decays and their implication for  supersymmetric flavor models,''
Nucl.\ Phys.\ B {\bf 540}, 37 (1999)
[hep-ph/9807377].
%%CITATION = HEP-PH 9807377;%%

\bibitem{HFAG}
We use the (ICHEP 2004) average from CLEO, BaBar and Belle 
compiled by the Heavy Flavor Averaging group. 
http://www.slac.stanford.edu/xorg/hfag/

%\cite{Grossman:1997dd}
\bibitem{Grossman:1997dd}
Y.~Grossman, Y.~Nir and M.~P.~Worah,
``A model independent construction of the unitarity triangle,''
Phys.\ Lett.\ B {\bf 407}, 307 (1997)
[hep-ph/9704287].
%%CITATION = HEP-PH 9704287;%%

\bibitem{BRANCOCP.BOOK}
G.~C.~Branco, L.~Lavoura and J.~P.~Silva,
``CP Violation,''
Clarendon Press, Oxford (1999).

%\cite{Anikeev:2001rk}
\bibitem{Anikeev:2001rk}
K.~Anikeev {\it et al.},
``B physics at the Tevatron: Run II and beyond,''
[hep-ph/0201071].
%%CITATION = HEP-PH 0201071;%%

%\cite{Buras:1998ra}
\bibitem{Buras:1998ra}
A.~J.~Buras,
``Weak Hamiltonian, CP violation and rare decays,''
[hep-ph/9806471].
%%CITATION = HEP-PH 9806471;%%

%\cite{Buras:1994dj}
\bibitem{Buras:1994dj}
A.~J.~Buras and M.~Munz,
``Effective Hamiltonian for B $\to$ X(s) e+ e- beyond leading logarithms in the NDR and HV schemes,''
Phys.\ Rev.\ D {\bf 52}, 186 (1995)
[hep-ph/9501281].
%%CITATION = HEP-PH 9501281;%%

%\cite{Buras:xp}
\bibitem{Buras:xp}
A.~J.~Buras, M.~Misiak, M.~Munz and S.~Pokorski,
``Theoretical Uncertainties And Phenomenological Aspects Of B $\to$ X(S) Gamma Decay,''
Nucl.\ Phys.\ B {\bf 424}, 374 (1994)
[hep-ph/9311345].
%%CITATION = HEP-PH 9311345;%%

%\cite{Chetyrkin:1996vx}
\bibitem{Chetyrkin:1996vx}
K.~G.~Chetyrkin, M.~Misiak and M.~Munz,
``Weak radiative B-meson decay beyond leading logarithms,''
Phys.\ Lett.\ B {\bf 400}, 206 (1997)
[Erratum-ibid.\ B {\bf 425}, 414 (1998)]
[hep-ph/9612313].
%%CITATION = HEP-PH 9612313;%%

%\cite{Grinstein:vj}
\bibitem{Grinstein:vj}
B.~Grinstein, R.~P.~Springer and M.~B.~Wise,
``Effective Hamiltonian For Weak Radiative B Meson Decay,''
Phys.\ Lett.\ B {\bf 202}, 138 (1988).
%%CITATION = PHLTA,B202,138;%%

%\cite{Demir:2001yz}
\bibitem{Demir:2001yz}
D.~A.~Demir and K.~A.~Olive,
``B $\to$ X/s gamma in supersymmetry with explicit CP violation,''
Phys.\ Rev.\ D {\bf 65}, 034007 (2002)
[hep-ph/0107329].
%%CITATION = HEP-PH 0107329;%%

%\cite{Barbieri:1993av}
\bibitem{Barbieri:1993av}
R.~Barbieri and G.~F.~Giudice,
``b $\to$ s gamma decay and supersymmetry,''
Phys.\ Lett.\ B {\bf 309}, 86 (1993)
[hep-ph/9303270].
%%CITATION = HEP-PH 9303270;%%

\bibitem{mybsgwk}
%\cite{Ali:1998rr}
%\bibitem{Ali:1998rr}
A.~Ali, H.~Asatrian and C.~Greub,
``Inclusive decay rate for B $\to$ X/d + gamma in next-to-leading  logarithmic order and CP asymmetry in the standard model,''
Phys.\ Lett.\ B {\bf 429}, 87 (1998)
[hep-ph/9803314];
%\cite{Kagan:1998ym}
%\bibitem{Kagan:1998ym}
A.~L.~Kagan and M.~Neubert,
``{QCD} anatomy of B $\to$ X/s gamma decays,''
Eur.\ Phys.\ J.\ C {\bf 7}, 5 (1999)
[hep-ph/9805303].
%%CITATION = HEP-PH 9805303;%%

%\cite{Kagan:1998bh}
\bibitem{Kagan:1998bh}
A.~L.~Kagan and M.~Neubert,
``Direct CP violation in B $\to$ X/s gamma decays as a signature of new  physics,''
Phys.\ Rev.\ D {\bf 58}, 094012 (1998)
[hep-ph/9803368].
%%CITATION = HEP-PH 9803368;%%

\bibitem{mybsgwkbr}
%\cite{Bertolini:1987pk}
%\bibitem{Bertolini:1987pk}
S.~Bertolini, F.~Borzumati and A.~Masiero,
``Supersymmetric Enhancement Of Noncharmed B Decays,''
Nucl.\ Phys.\ B {\bf 294}, 321 (1987);
%%CITATION = NUPHA,B294,321;%%
%\cite{Barbieri:1993av}
%\bibitem{Barbieri:1993av}
R.~Barbieri and G.~F.~Giudice,
``b $\to$ s gamma decay and supersymmetry,''
Phys.\ Lett.\ B {\bf 309}, 86 (1993)
[hep-ph/9303270];
%%CITATION = HEP-PH 9303270;%%
%\cite{Cho:1996we}
%\bibitem{Cho:1996we}
P.~L.~Cho, M.~Misiak and D.~Wyler,
``$K_L \to \pi~0 e~+ e~-$ and $B \to X_s \ell~+ \ell~-$ Decay in the MSSM,''
Phys.\ Rev.\ D {\bf 54}, 3329 (1996)
[hep-ph/9601360].
%%CITATION = HEP-PH 9601360;%%
%\cite{Hewett:1996ct}
%\bibitem{Hewett:1996ct}
J.~L.~Hewett and J.~D.~Wells,
``Searching for supersymmetry in rare B decays,''
Phys.\ Rev.\ D {\bf 55}, 5549 (1997)
[hep-ph/9610323].
%%CITATION = HEP-PH 9610323;%%

%\cite{Nishida:2003yw}
\bibitem{Nishida:2003yw}
S.~Nishida {\it et al.}  [BELLE Collaboration],
``Measurement of the CP asymmetry in B $\to$ X/s gamma,''
Phys.\ Rev.\ Lett.\  {\bf 93}, 031803 (2004)
[hep-ex/0308038].

%%CITATION = HEP-EX 0308038;%%    
%\cite{Aubert:2004hq}
\bibitem{Aubert:2004hq}
B.~Aubert {\it et al.}  [BABAR Collaboration],
``Measurement of the direct CP asymmetry in b $\to$ s gamma decays,''
Phys.\ Rev.\ Lett.\  {\bf 93}, 021804 (2004)
[hep-ex/0403035].
%%CITATION = HEP-EX 0403035;%%

%\cite{Kagan:1997qn}
\bibitem{Kagan:1997qn}
A.~L.~Kagan and J.~Rathsman,
``Hints for enhanced b $\to$ s g from charm and kaon counting,''
[hep-ph/9701300].
%%CITATION = HEP-PH 9701300;%%

%\cite{Kaneko:2002mr}
\bibitem{Kaneko:2002mr}
J.~Kaneko {\it et al.}  [Belle Collaboration],
``Measurement of the electroweak penguin process B $\to$ X/s l+ l-,''
Phys.\ Rev.\ Lett.\  {\bf 90}, 021801 (2003)
[hep-ex/0208029].
%%CITATION = HEP-EX 0208029;%%

%\cite{Beneke:2001ev}
\bibitem{Beneke:2001ev}
M.~Beneke, G.~Buchalla, M.~Neubert and C.~T.~Sachrajda,
``QCD factorization in B $\to$ pi K, pi pi decays and extraction of  Wolfenstein parameters,''
Nucl.\ Phys.\ B {\bf 606}, 245 (2001)
[hep-ph/0104110].
%%CITATION = HEP-PH 0104110;%%

%\cite{Beneke:2003zv}
\bibitem{Beneke:2003zv}
M.~Beneke and M.~Neubert,
``QCD factorization for B $\to$ P P and B $\to$ P V decays,''
Nucl.\ Phys.\ B {\bf 675}, 333 (2003)
[hep-ph/0308039].
%%CITATION = HEP-PH 0308039;%%

%\cite{Aubert:2002nx}
\bibitem{Aubert:2002nx}
B.~Aubert {\it et al.}  [BABAR Collaboration],
``Measurement of sin(2beta) in B0 $\to$ Phi K0(S). ((B)),''
hep-ex/0207070.
%%CITATION = HEP-EX 0207070;%%

%\cite{Abe:2002bx}
\bibitem{Abe:2002bx}
K.~Abe {\it et al.}  [Belle Collaboration],
``An improved measurement of mixing-induced CP violation in the neutral B  meson system,''
hep-ex/0207098.
%%CITATION = HEP-EX 0207098;%%

%\cite{Hahn:1998yk}
\bibitem{Hahn:1998yk}
T.~Hahn and M.~Perez-Victoria,
``Automatized one-loop calculations in four and D dimensions,''
Comput.\ Phys.\ Commun.\  {\bf 118}, 153 (1999)
[hep-ph/9807565].
%%CITATION = HEP-PH 9807565;%%

\end{thebibliography}
\end{document}